# PARIS BETA

*Study of Requirements and Mission Definition for Bistatic Altimetry*

Starlab, Astrium, CLS, GMV, IEEC, Ifremer
ESA/ESTEC Contract No. 15083/01/NL/MM

# Mesoscale Ocean Altimetry Requirements and Impact of GPS-R measurements for Ocean Mesoscale Circulation Mapping

Technical Note Extract from the PARIS-BETA ESTEC/ESA Study
**ESTEC Technical Officer: Pierluigi Silvestrin**

Written by
***P.Y. Le Traon and G. Dibarboure (CLS), G. Ruffini (Starlab), E. Cardellach (IEEC)***

Contact: Giulio@starlab-bcn.com

## CONTENTS

# 1 Introduction: the PARIS BETA study

PARIS-Beta, a Starlab lead ESA project with Astrium, CLS, GMV, IEEC, Ifremer as sub-contractors (2001-2002), is part of the ongoing European (ESA) effort to demonstrate the potential for a novel class of Earth Observation measurements based on GNSS-R. GNSS-R is a highly innovative EO technology, which can provide on a long-term basis much needed and unique data products with excellent coverage and resolution and at low cost. The underlying principle of this technology is the use of reflected signals originating from sources of opportunity to infer properties of the ocean or other reflecting surfaces. GNSS-R is a multi-static radar system in which only the receivers need to be deployed---the emitters already operating for other purposes.

In the framework of the PARIS Beta project, fundamental milestones have been reached for the definition of future GNSS-R (Global Navigation Satellite System signal Reflections) altimetry missions (the PARIS concept). The most important one is the confirmation of the significant impact that GNSS-R data can have on mesoscale oceanography, as we discuss here.

PARIS-Beta has performed a feasibility analysis of spaceborne GNSS-R altimetry. GNSS-R altimetry (PARIS) hinges on the availability of code and phase pseudo-range measurements from both direct and reflected GNSS signals (the first provide the foundation for GNSS navigation). The project sought to carefully define the user requirements and confront them with technology limitations. This process included a careful analysis of technology feasibility, simulation of GNSS-R altimetric data, algorithm development and impact studies both through simple averaging of the data and through assimilation into an advanced mesoscale ocean circulation model--both with very encouraging results. The study also included preliminary mission analysis, instrument concept and ground processing work packages. An important achievement of the study has been the development of a (raw data) Level 0 (ocean surface) GNSS-R simulator and the development of a signal processor for GNSS-R space applications, enabling the study of the characteristics of the reflected signal. This project has also benefited (and vice versa) from the results of the parallel project OPPSCAT 2, dedicated to understanding GNSS-R reflectometry applications

Sea level measurements are an essential component of a global ocean observing system. They are also crucial for most of the operational applications of oceanography (e.g. offshore and coastal applications, marine safety, Navies applications). Sea level measurements as derived from GPS signals may complement the existing of future systems based on satellite altimetry (TOPEX/POSEIDON, ERS-1/2 and later on Jason-1 and ENVISAT). They can, in particular, significantly improve the space/time sampling of the ocean, which will allow a better description of the ocean mesoscale circulation. This is crucial for the operational applications of oceanography as well as for a better description of the ocean for climate studies.

In this report, we first briefly review the contribution of satellite altimetry to the mesoscale oceanography. We then summarise recent results obtained on the mapping capabilities of existing and future altimeter missions. From these analyses, refined requirements for mesoscale ocean altimetry (in terms of space/time sampling and accuracy) are derived. A review of on-going and planned altimetric missions is then performed and we analyse how these configurations match the user requirements. Then we will describe the simulation approach and impact analysis of GPS-R data.

## 2 What did we learn from altimetry?

Satellite altimetry has made a unique contribution to observing and understanding mesoscale variability (see Le Traon and Morrow, 2000 for a recent review). Altimeter data analyses have provided, for the first time, a global description of the eddy energy and its seasonal/interannual variations. The time and space scales of the mesoscale circulation have been characterized. The eddy/mean flow interactions have also been mapped and provide an important ingredient for understanding the western boundary current and ACC dynamics. Satellite altimetry has also allowed a synoptic mapping of large eddies (e.g. Agulhas eddies) which is useful to better understand the role of eddies in transporting mass, heat, salt and nutrients. All these studies provide a good means of testing and validating models and theories.

Most of these studies have used GEOSAT, T/P or ERS altimeter data separately. The contribution of the merging of T/P and ERS is, however, well illustrated by Ducet et al. (2000) and Ducet and Le Traon (2001). The sea level and the velocity fields were mapped globally; this yielded a characterization of the Eddy Kinetic Energy (EKE), anisotropy and eddy mean flow interactions with a resolution never achieved before (see figures 1 and 2).

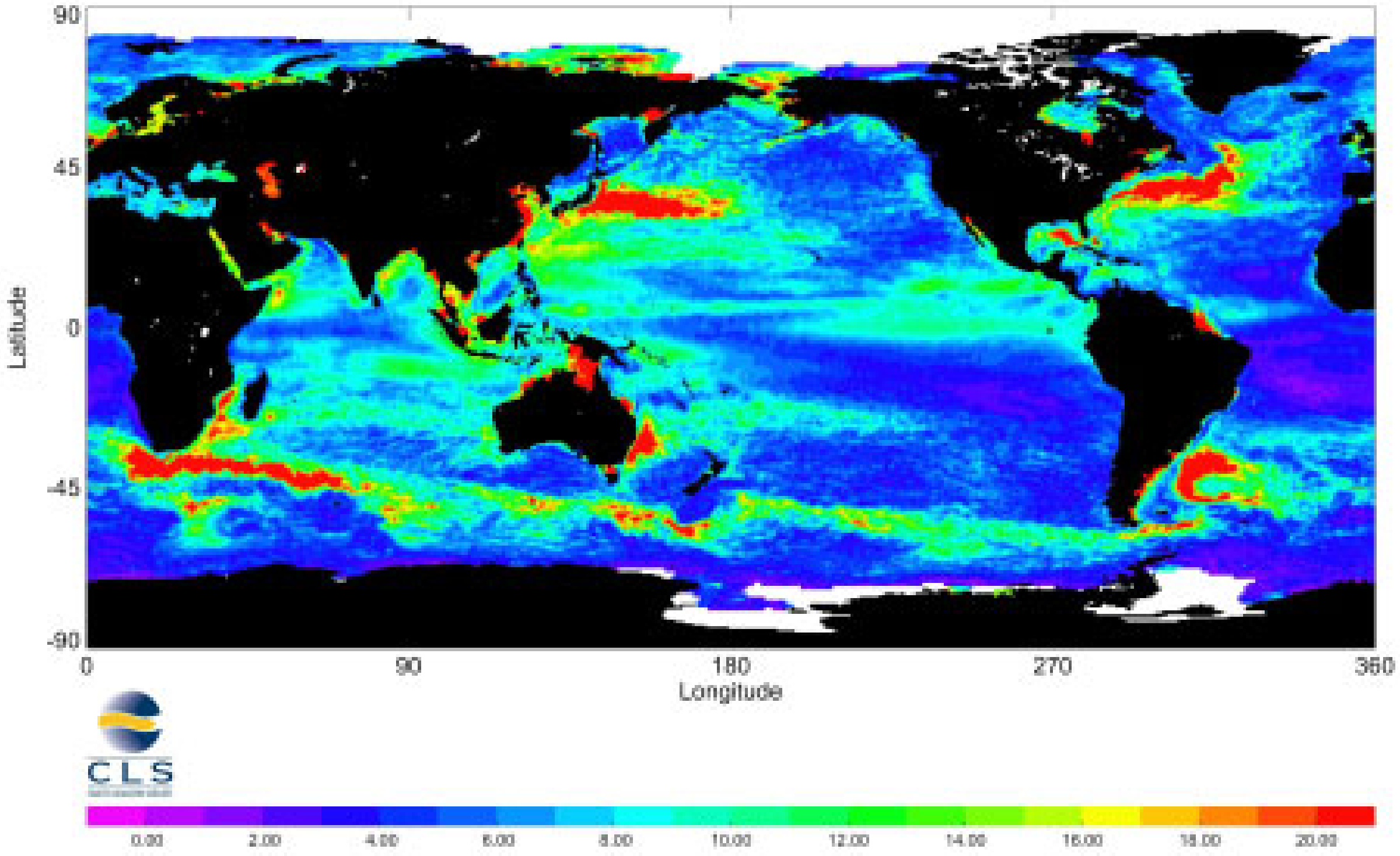

Figure 1: Rms of sea level variations derived from the combination of T/P and ERS-1/2 over a 5 year period (Ducet et al., 2000). The map presents a very detailed description presumably never before achieved at a global scale. Units are cm.

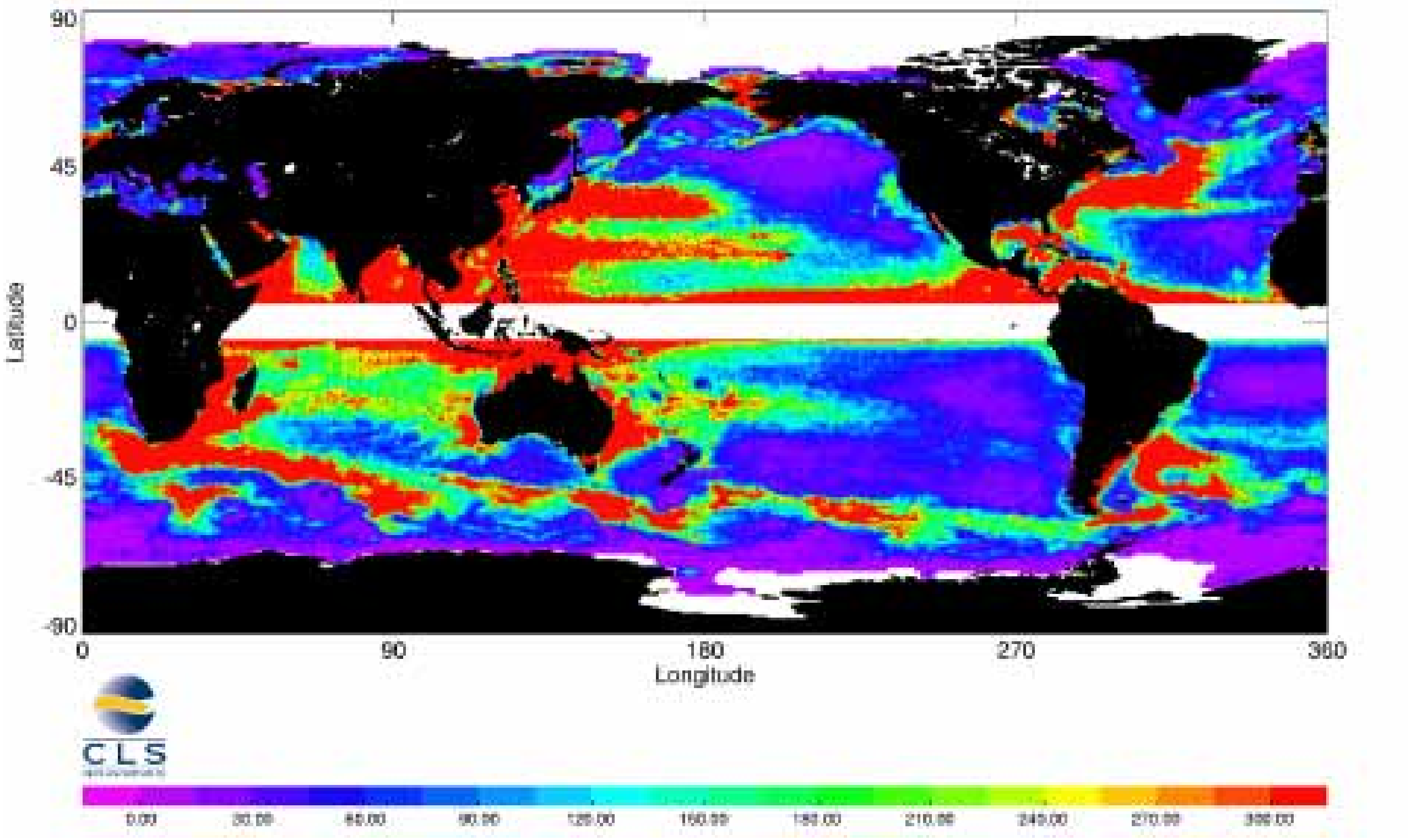

Figure 2: Eddy Kinetic Energy derived from the combination of T/P and ERS-1/2 over a 5 year period (Ducet et al., 2000). Units are $cm^2/s^2$. (ranging from 0-300).

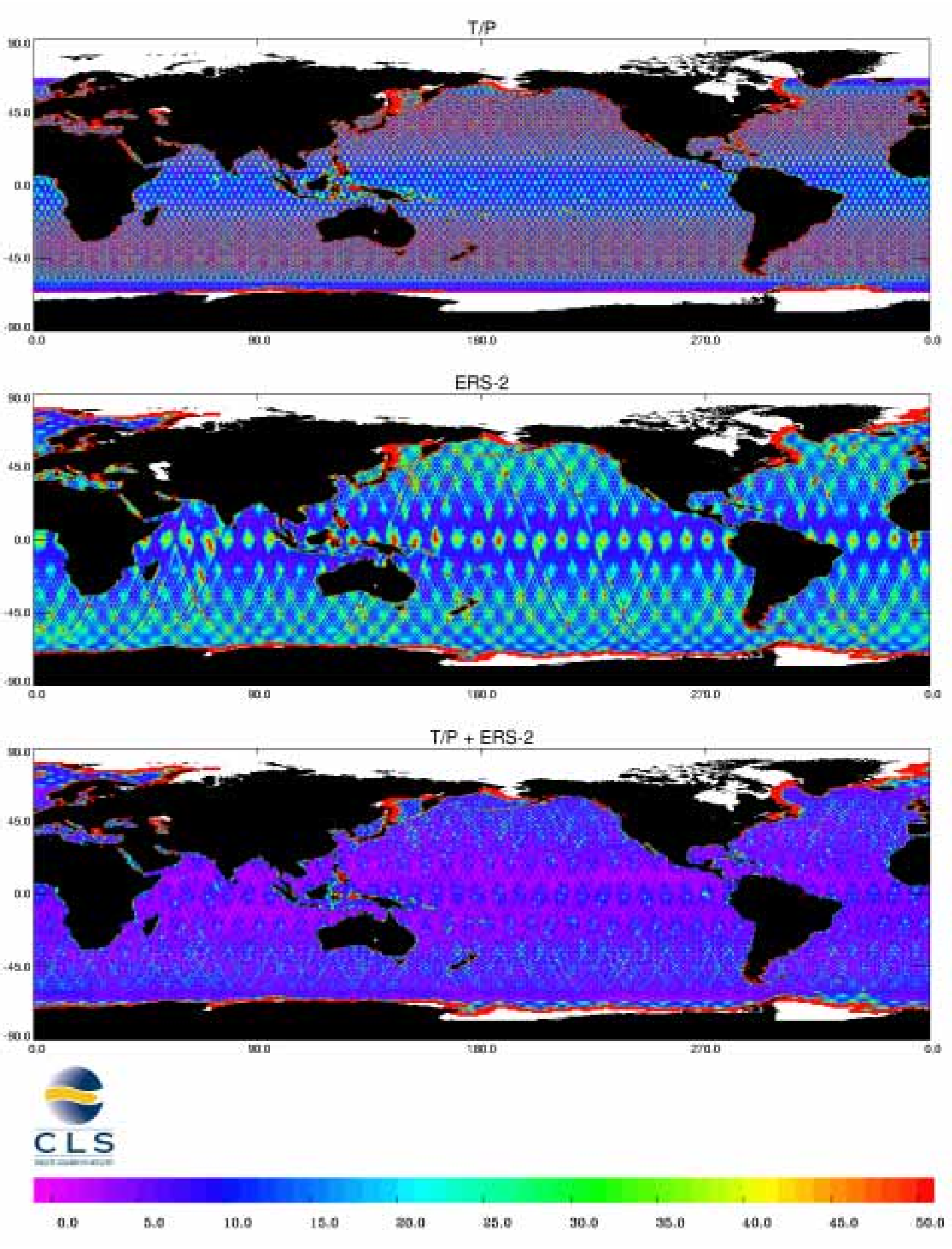

Figure 3: Formal sea level mapping error (in percentage of signal variance) for T/P, ERS and the combination of T/P and ERS (Ducet et al., 2000). Realistic space and time scales of mesoscale variability are taken into account in the calculation. [scale ranges from 0-50].

Ducet et al. (2000) showed that the sea level can be mapped with an accuracy of about 10-20% of the signal variance (see figure 3) and the velocity field with an accuracy varying from 20% to 40% of the signal variance from low to high latitudes. These results were derived from the comparison of maps (and mapping errors) derived from T/P and ERS alone and from the comparison with WOCE drifter data (see Ducet et al., 2000 for a detailed discussion). They are consistent with those derived from formal error analyses (see below).

# 3 Requirements for sea level measurements

## 3.1 Main requirements for climate and mesoscale applications

The usually agreed main requirement for future altimeter missions is that at least two (and preferably three) altimeter missions with one very precise long-term altimeter system are needed (e.g. Koblinsky et al., 1992). The long-term altimeter system is supposed to provide the low frequency and large scale climatic signals and to provide a reference for the other altimeter missions. It requires a series of very precise (centimetre level) and inter-calibrated missions. TOPEX/POSEIDON and later on the Jason series have been designed to meet these objectives. The role of the other missions is to provide the higher wavenumbers and frequencies and, in particular, the mesoscale signal, which cannot be well observed with a single altimeter mission. This does not require precise altimeter systems as most of the altimetric errors (in particular the orbit error) are at long wavelengths and they do not impact significantly the mesoscale signal.

Such a requirement for future altimeter missions is partly based on several studies on the sampling characteristics of single and multiple altimeter missions. Le Traon and Dibarboure (1999) (hereafter LD99) and Le Traon et al. (2001) (hereafter LDD01) have, in particular, quantified the contribution of single and multiple altimeter missions for the mapping of mesoscale variability. In the following section, we summarise the main findings of these studies as well as some new results we obtained more recently. These studies are then used to provide refined requirements for mesoscale applications.

## 3.2 Summary of LDD99 and LDD01 studies

LD99 have quantified the mesoscale mapping capability when combining various existing or future altimeter missions in terms of sea level anomaly (SLA) (figure 4) and zonal (U) and meridional (V) velocity. Their main results are as follows:

- The GEOSAT (or GEOSAT Follow On) 17-day orbit provides the best sea level and velocity mapping for the single-satellite case. The T/P+Jason-1 (interleaved T/P - Jason-1 tandem orbit scenario) provides the best mapping for the two-satellite case. There is only minor improvement, however, with respect to the T/P+ERS (or Jason-1+ENVISAT) scenario.

- There is a large improvement in sea level mapping when two satellites are included. For example, compared to T/P alone, the combination of T/P and ERS has a mean mapping error reduced by a factor of 4 and a standard deviation reduced by a factor of 5.

- The velocity field mapping is more demanding in terms of sampling. The U and V mean mapping errors are two to four times larger than the SLA mapping error. Only a combination of three satellites can provide a velocity field mapping error below 10% of the signal variance.

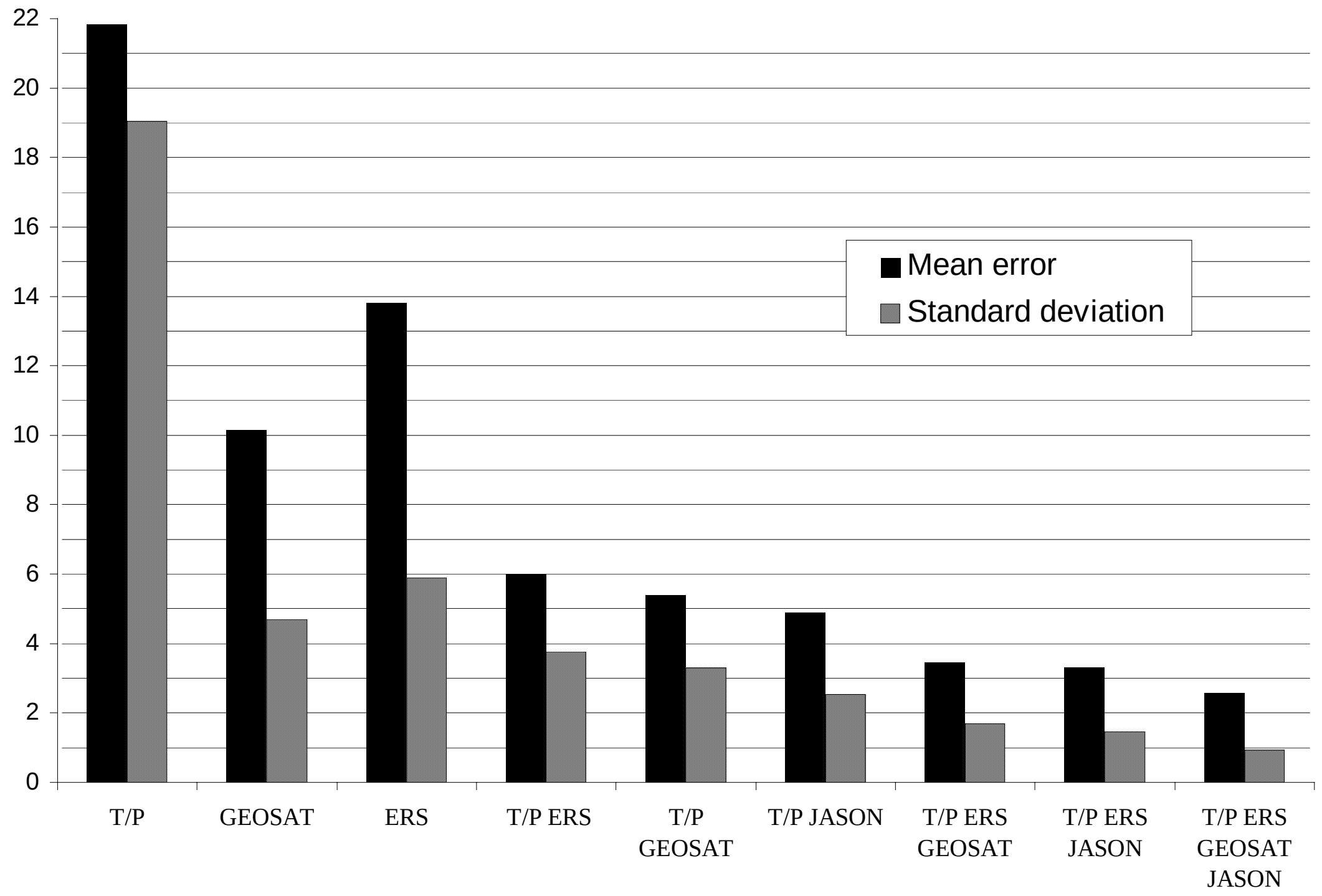

Figure 4 : Mean and standard deviation of Sea Level Anomaly (SLA) mapping error for single and multiple altimeter missions. Units are in % of signal variance. The calculation assumes a space scale of 150 km, a time scale of 15 days and a noise/signal ratio of 2%.

The LD99 study was extended by LDD01 who analysed the sea level mapping capabilities of multiple altimeters using the Los Alamos North Atlantic high-resolution model (Smith et al., 2000). Los Alamos Model (hereafter LAM) represents the mesoscale variability quite well and offered a unique opportunity for assessing the mapping capability of multiple altimeter missions.

LDD01 have shown that sea level mapping errors were larger than the ones derived from LD99 formal error analysis (by a factor of 1.5 to 2). This was mainly due to high frequency signals. In areas with large mesoscale variability, these signals represent 5 to 10% of the total sea level variance (figure 5a) (see also Minster and Gennero, 1995) and are associated with high wavenumbers. They account for 15 to 20 % of the total velocity variance (figure 5b). In shallow and high latitude regions, these high frequency signals account for up to 30-40% of the total sea level and velocity variance; there, part of these signals correspond to large scale barotropic motions.

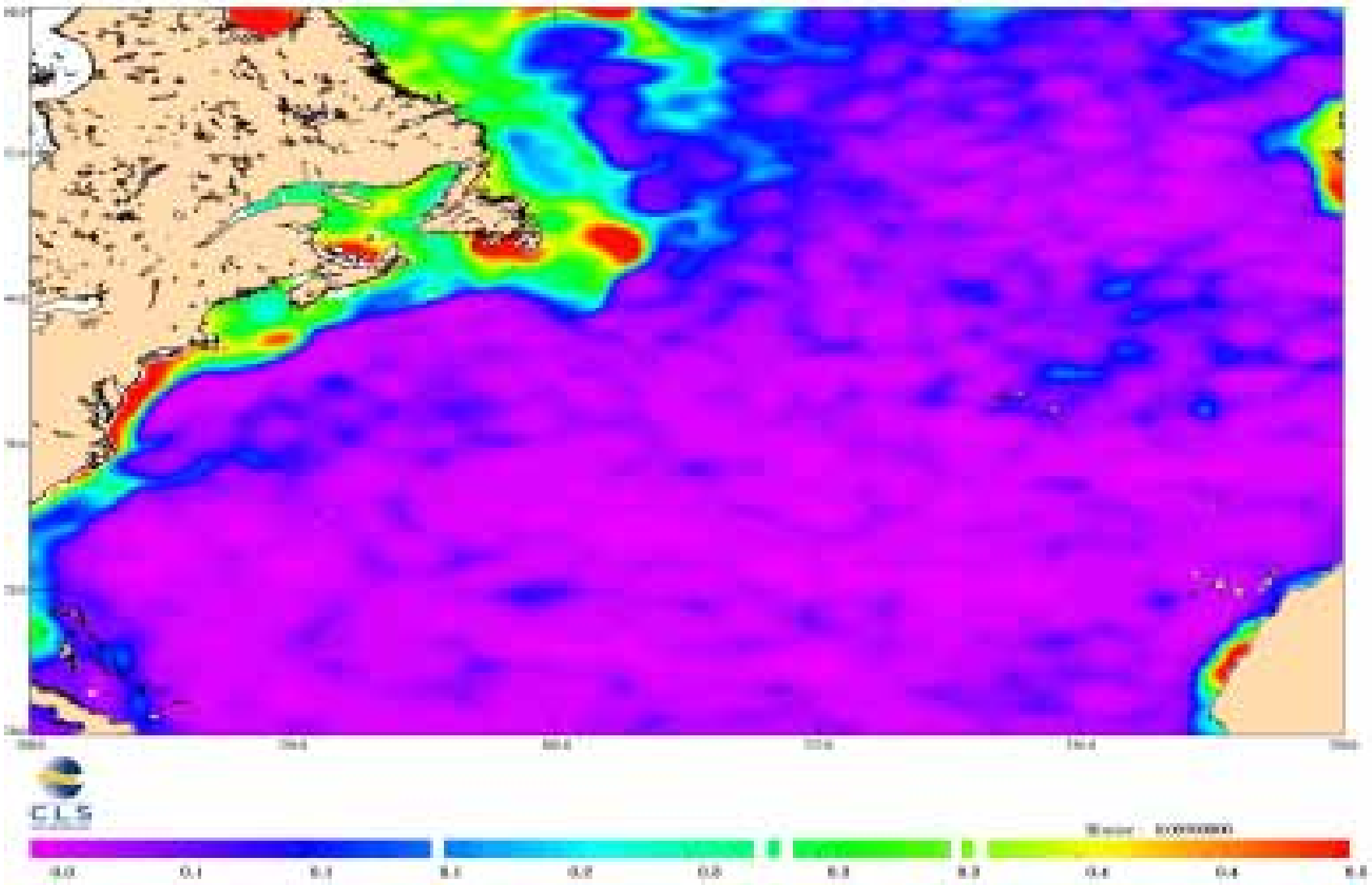

Figure 5a: Contribution of high frequency signals (periods < 20 days) to the total sea level variance for the Los Alamos simulation.

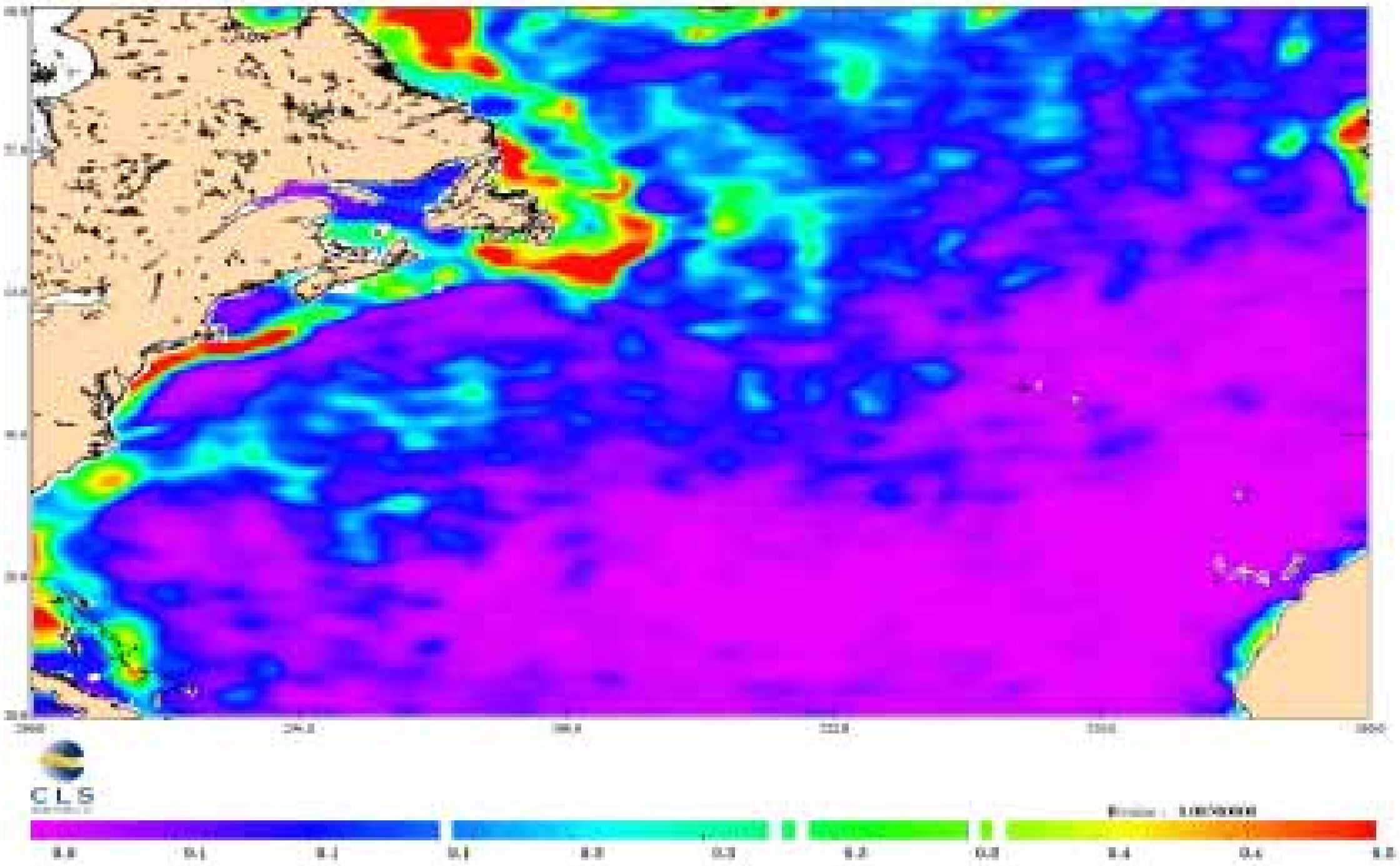

Figure 5b: Contribution of high frequency signals (periods < 20 days) to the total zonal velocity variance for the Los Alamos simulation. [scale ranges from 0-0.5].

LDD01 study was recently extended to the analysis of velocity field mapping capabilities. We also analysed “optimised” three and four satellites configurations (three and four interleaved Jason-1). To better analyse the impact of the high frequency signals on the sea level and velocity mapping, we systematically computed for all analysed configurations the mapping errors on the instantaneous fields (as in LDD01) and on 10-day averaged fields. For the latter, the estimated fields were compared to the model fields filtered using a Loess filter with a cut-off period of 16 days (which chiefly corresponds to a 10-day average). Results for the T/P+ERS configuration are shown on figures 6a (sea level) and 6b (zonal velocity). Results for the different configurations in the Gulf Stream area are summarised in table 1.

| | H | U | V |
|---|---|---|---|
| ***T/P + ERS (Jason-1 + ENVISAT)*** | 8 / 4.9 | 29 / 10.9 | 40 / 15.4 |
| T/P+ERS+Jason-1 | 6.5 / 2.6 | 23 / 6.9 | 29 / 9.7 |
| Three interleaved Jason-1 | 4.8 / 1.9 | 20.2 / 5.6 | 21.9 / 6.5 |
| Four interleaved Jason-1 | 4.3 / 1.7 | 18.9 / 5.1 | 20.8 / 5.4 |

Table 1: Sea Level (H), zonal (U) and meridional (V) velocity mean mapping error over the Gulf Stream area (34°N-39°N – 70°W-60°W) as derived from the simulation using the Los Alamos model fields. Errors expressed in percentage of the total sea level and velocity variance are given both for “instantaneous” and 10-day average signal mapping.

Compared to T/P+ERS, a three or four “optimised” satellite configuration will allow a large improvement in the description of mesoscale variability. However, to achieve a mapping accuracy of the total velocity field better than 15-20% of the signal variance, one needs to resolve the high frequency and high wavenumber signals. This will require a much denser space and time sampling (better than 100 km and 10 days). The aliasing of the high frequency signals is also an important issue; thus, even if these signals cannot be resolved, one needs to take into account their impact on lower frequencies/wavenumbers. Note that the large-scale high frequency barotropic signals (e.g. Tierney et al., 2000) are likely to be well resolved by an “optimised” three or four satellite configuration.

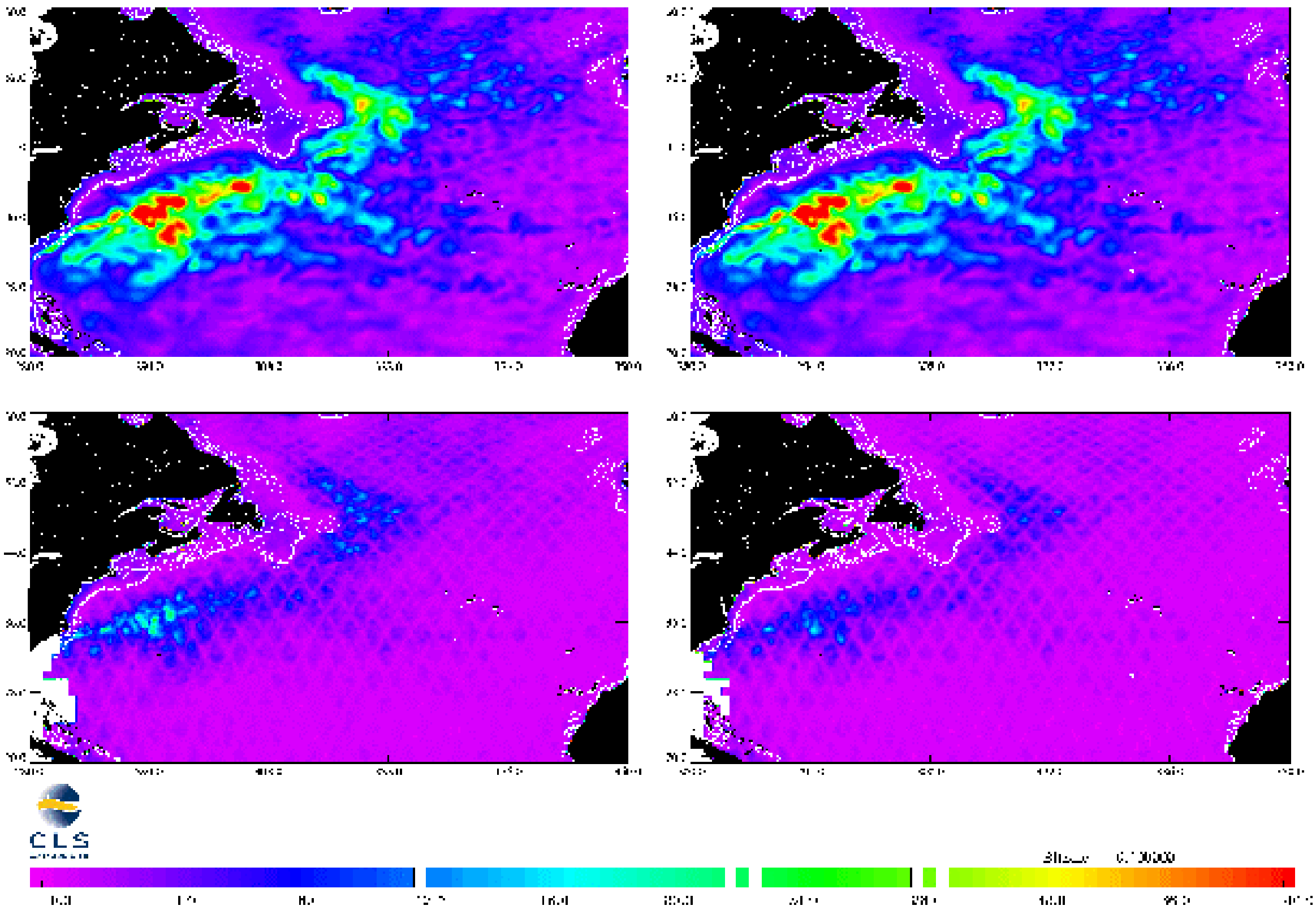

Figure 6a : Rms sea level for the Los Alamos Model simulation (upper figures). Rms sea level mapping error for T/P+ERS (lower figures). Left panel corresponds to instantaneous signal mapping error. Right panel corresponds to 10-day average signal mapping error. Units are cm.

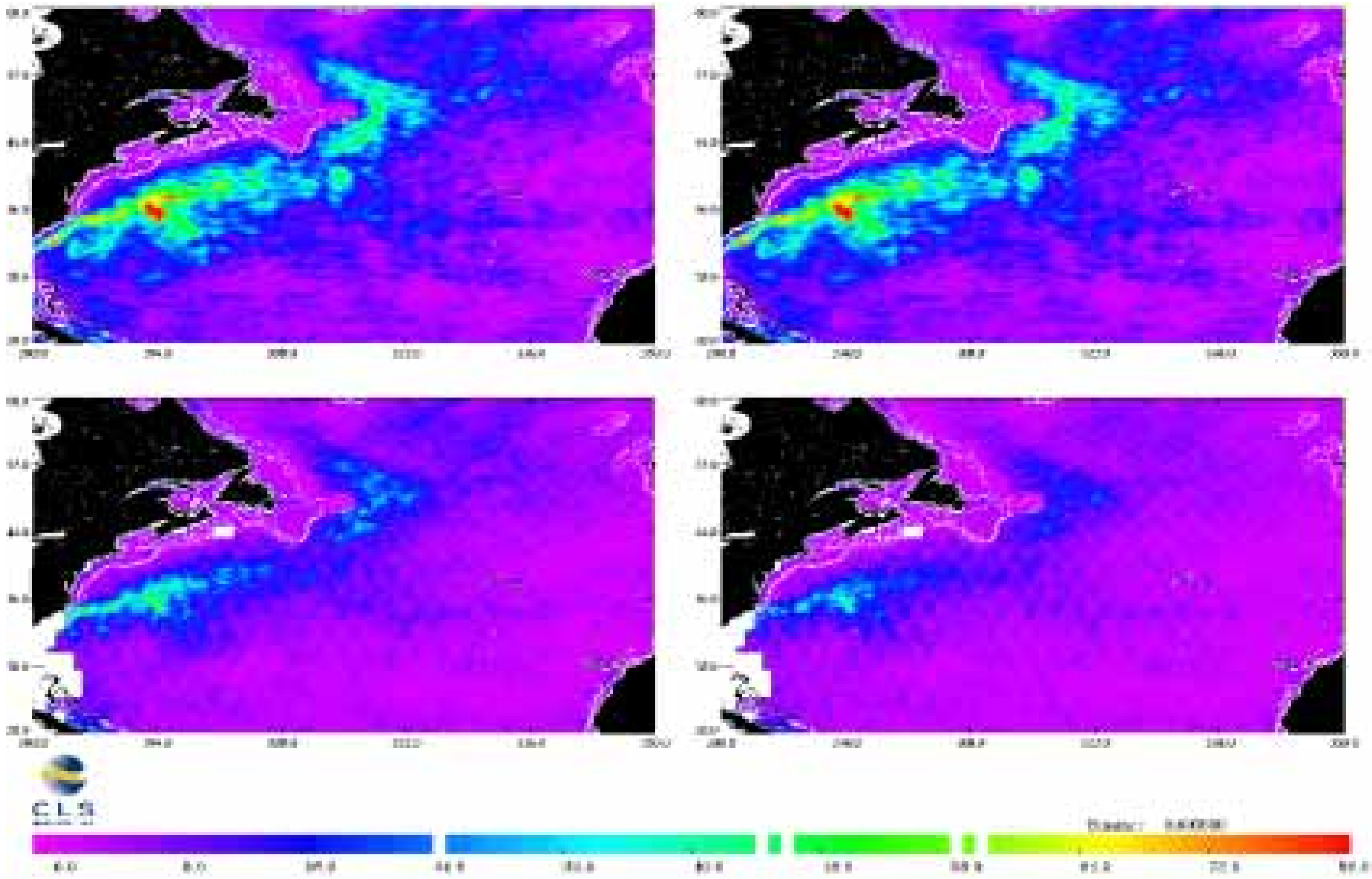

Figure 6b : Rms zonal velocity for the Los Alamos Model simulation (upper figures). Rms zonal velocity mapping error for T/P+ERS (lower figures). Left panel correspond to instantaneous signal mapping error. Right panel correspond to 10-day average signal mapping error. Units are cm/s. [scale from 0-80].

## 3.3 Refined requirements for mesoscale applications

From these studies, we can derive refined the following refined requirements for mesoscale applications:

1. The minimum requirement will be to continue flying a two-satellite configuration (after Jason-1 and ENVISAT). It provides already a good representation of the mesoscale variability (sea level mapping errors of the order of 10% of the signal variance). It cannot provide, however, a sufficiently accurate estimation of the velocity field (e.g. below 10% of the signal variance) and will not allow us to track small eddies (e.g. diameter below 100 km).
2. This can be significantly improved with an optimised three-satellite configuration. Compared to Jason-1 and ENVISAT (or T/P+ERS), such a configuration should allow a reduction of sea level and velocity mapping errors by a factor of about 2 to 3. It is also likely to partly resolve the large-scale high frequency barotropic motions.
3. To further improve the mapping (which would be needed for some of the envisioned scientific and operational applications – see section 3.5), we need to resolve the high frequency and high wavenumber signals, i.e. sample the ocean with a time sampling below 10 days and 100 km. This is likely to require constellation of up to six satellites and/or use different concepts for satellite altimetry.

This analysis only deals with sampling requirements. As far as measurement errors are concerned, the following requirements can be made:

1. Assuming the Jason series continue to provide a long-term reference, the additional measurement systems do not have to provide very precise measurements. Results derived from these systems will not be sensitive to long wavelengths errors (wavelengths > 1000-2000 km) if the Jason satellites are used to constrain the large-scale (climatic) signals.
2. Typical amplitude of mesoscale signal is 4 to 8 cm rms in the open ocean and 20 to 40 cm rms in the high eddy energy regions (see figure 1, noting that in low eddy energy regions the mesoscale variability accounts for approximately half of the total sea level variability signal). A 3 cm measurement noise is thus satisfactory but a smaller noise will allow a better estimation of the velocity fields (i.e., sea level gradients) and a detailed analysis of the eddy structure. A larger measurement noise will still be useful, however, for mesoscale applications. . For a single observation with a noise variance of $\varepsilon^2$, the relative mapping error (in %) at the observation location is equal to $100(1-Var/(Var+\varepsilon^2))$ where Var is the signal variance. For example, in low eddy energy regions, an observation with a 6 cm rms measurement noise will yield a relative mapping error of 50% in low eddy energy regions and 4% in high eddy energy regions. Given that the mapping errors with T/P or Jason-1 can reach up to 70% of the signal variance between tracks, such an observation will provide a useful additional information everywhere if the Jason series is not complemented by other satellites. With Jason-1 and ENVISAT, the mapping error will be always below 20% of the signal variance but the additional information will still very significantly reduce the mapping error in high eddy energy regions.

For GPS signals, we will also have to deal with geoid errors, as measurements will not be obtained along exact repeat tracks. This additional source of error has to be carefully quantified (taking into account the improvement of geoid models due to altimetry and the CHAMP/GRACE/GOCE gravimetric missions). Other sources of errors will also have to be precisely quantified (e.g. propagation effects). Of particular importance is to establish the error spectrum and, in particular, the degree of error correlation between individual measurements.

## 3.4 The role of data assimilation

One should note that the best use of high resolution altimetry data will be when they are assimilated with in-situ and other remote sensing data into global eddy resolving models (GODAE) (and nested shelf/coastal models). This will open a large range of scientific and operational applications (see section 3.5). Data assimilation is a powerful means for a dynamic interpolation of data and may alter the spatial and temporal requirements (e.g. using a model forecast as an a priori knowledge instead of Climatology). Demonstrating the value of data assimilation is the central objective of GODAE, the Global Ocean Data Assimilation Experiment (Smith and Lefebvre, 1997; Le Traon et al., 1999). Providing general requirements from a data assimilation perspective is, however, today a very difficult issue. Results are not always consistent and are highly dependant on the data assimilation method and the model used to assess the contribution of altimeter data. Although data assimilation techniques should be ultimately used to determine the impact of altimeter data, they are not mature enough (at least for global eddy resolving models) to provide sufficiently general requirements.

The requirements presented in the previous sections thus do not take into account the model contribution as a dynamic interpolator of spare measurements. They take, however, into account an a priori knowledge of the space and time scales of the mesoscale variability as well as the noise characteristics. They can be considered as a robust estimation of the contribution of the data themselves.

## 3.5 Contribution of high resolution altimetry

Despite all the progress made in the last decade, there is still much to learn from altimeter data for mesoscale variability studies. Investigations that could be carried out include:

- More detailed comparisons of altimetry (including comparison of higher order statistics such as frequency/wavenumber spectra and Reynolds stresses) with eddy resolving models.
- Regional characterization of the 3D frequency/wavenumber of sea level (and velocity) and relation with forcings and dynamics. Relation with turbulence theories.
- Better characterization of seasonal/interannual variations in eddy energy and relation with forcings (mean current instabilities, winds)

- Phenomenological (global) characterization of eddies (eddy census) : size, rotation, diameter, life time, propagation….Relation with theories and models.
- Detailed dynamical structure of eddies. Estimation of the vorticity field (in and out of the eddy), divergence and deformation fields. Use in synergy with high resolution SST and Ocean Colour images. Estimation of vertical circulation and biogeochemical coupling.
- Relation and interaction between eddies and Rossby waves.
- Eddy heat fluxes (in combination with SST remote sensing data). Contribution to the total heat fluxes.
- Eddy mean flow interaction. Role of eddies on the general circulation and climate.

All these studies will benefit from higher space and time resolution. In particular, the phenomenological and dynamical characterisation of eddies and the eddy mean flow interaction studies will require a much higher resolution than the one we have today (i.e. with two satellites).

Finally note that a higher space and time resolution will allow a much more precise estimation of the velocity field that will be of great importance for most of the operational applications of satellite altimetry (offshore, fisheries, marine safety,…).

# 4 Review of on-going and planned altimetric missions

Table 2 below shows the existing and future (i.e. almost decided) altimetric missions (courtesy A. Ratier). As can be seen, the post-Jason-1 scenario is far from being satisfactory, as a follow on mission for ENVISAT is not yet planned. The minimum requirement for mesoscale oceanography is not yet guaranteed. In addition as explained above, there is much to learn on mesoscale oceanography with a better space/time sampling. This opens up interesting perspective for potential new systems (such as GPS altimetry) if the sampling and noise characteristics are satisfactory. This will be analysed in the second part of this study where the techniques used by LD99 and LDD01 (formal error analysis and simulations with the Los Alamos model) will be applied to quantify the contribution of GPS signals.

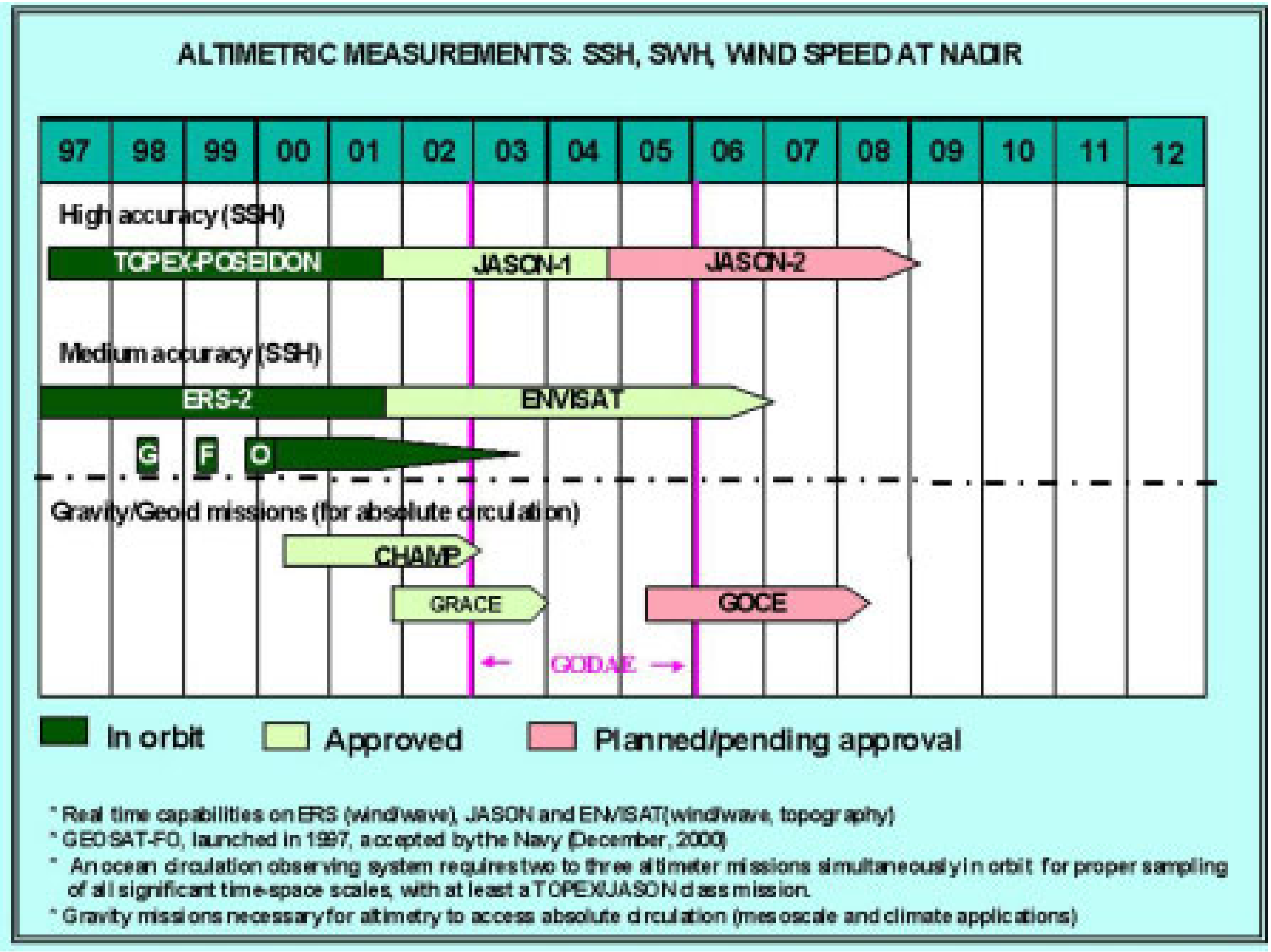

Table 2 : Planning of existing and future altimeter missions (courtesy A. Ratier).

# 5 GPS-R Impact Study: Introduction

In this section, the contribution of GPS altimetry for the mapping of ocean mesoscale circulation is quantified using the Los Alamos North Atlantic model. The Los Alamos model is known to represent the mesoscale variability quite well and is very well suited to simulate the contribution of sea level measurements (Le Traon et al., 2001). Model sea level fields are sub sampled to simulate the typical space/time sampling of GPS altimeter systems. A realistic measurement noise is added to these simulated measurements. We then evaluate how these simulated measurements can be used to reconstruct the initial model reference fields depending on the different regions and dynamical regimes. This is achieved using a space/time objective mapping technique that takes into account the GPS measurement noise characteristics and an a priori information on the space and time scales of ocean signals. Formal sea level mapping errors (see Le Traon and Dibarboure, 1999) are also analysed as they provide an estimation of mapping errors that depend only on sampling, noise and ocean signal characteristics.

# 6 Simulation of the GPS sampling and noise characteristics

A typical space/time sampling of multi-beam GPS altimetry was used. It corresponds to the GPS reflections at the Earth surface from the September 15, 2001 GPS constellation that would be received by a single LEO satellite in a 500 km polar orbit. Given the chosen LEO orbit, this sampling repeats approximately every three days. For this simulation, we only selected the best six reflections (six beam limitation). Figure 1a shows the corresponding sampling over a 3-day period over the whole North Atlantic. Colours correspond to different incident angles.

The measurement errors $\varepsilon_{GPS}$ for 1 second sampling were taken as:

$$\varepsilon_{GPS} = 30 \text{ cm} /(2 \cos \theta) \qquad (1)$$

where $\theta$ is the incidence angle (i.e., $\theta = 0$ for nadir observations). This corresponds to a PETREL-like mission (with a multi-beam upgrade), i.e. a 500 km orbit, antenna gain of 25-28 dB, no dual polarization, complex sampling and one bit quantization. In practice, to reduce the number of data points in the mapping procedure, a 3-second sampling was used (i.e. approximately every 21 km along the pseudo-tracks) and the errors given in (1) were divided by $\sqrt{3}$. These errors are represented on Figure 1b. This yields, for example, an error of 12 cm rms for a 45° incidence angle.

a)

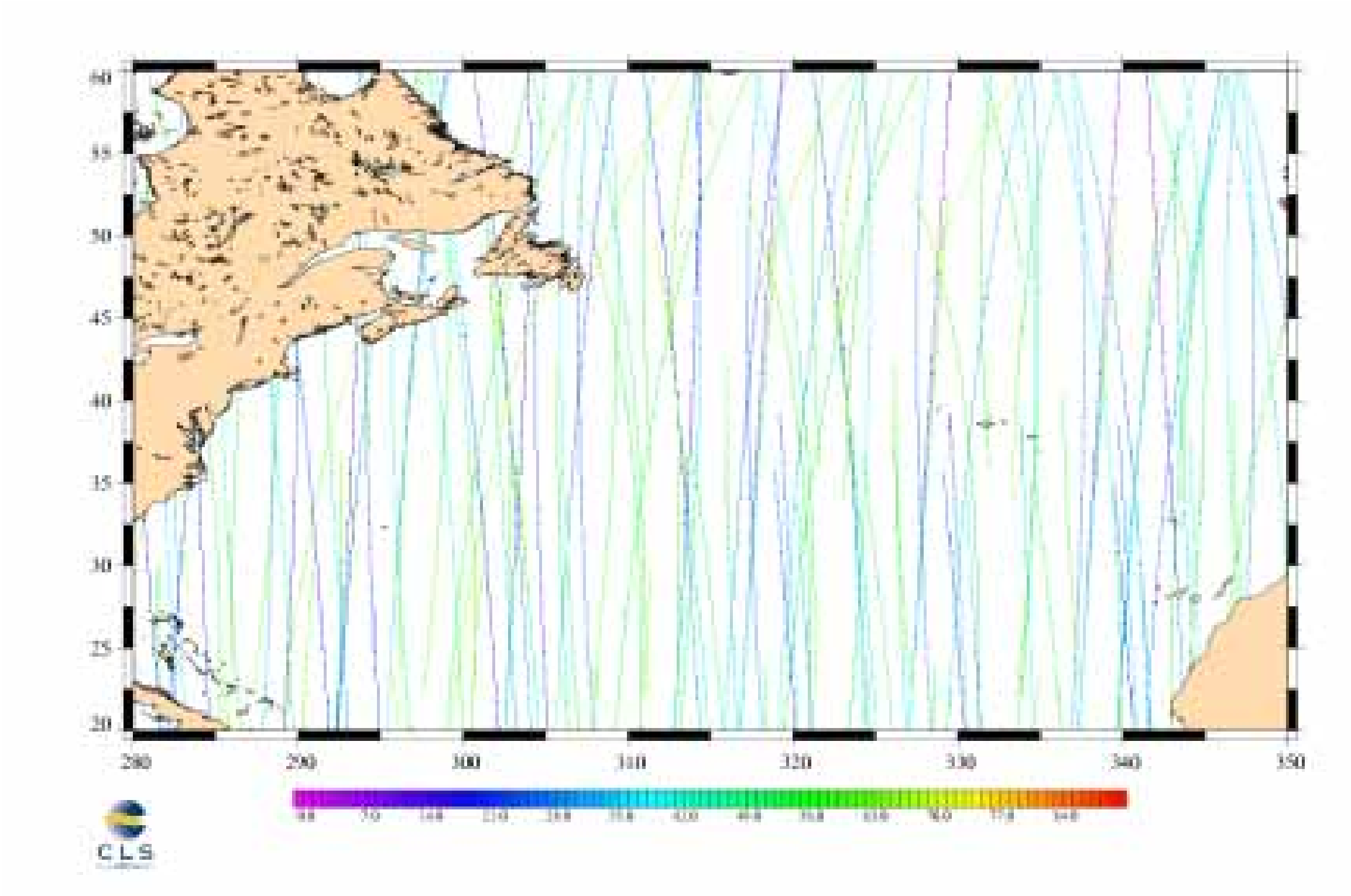

b)

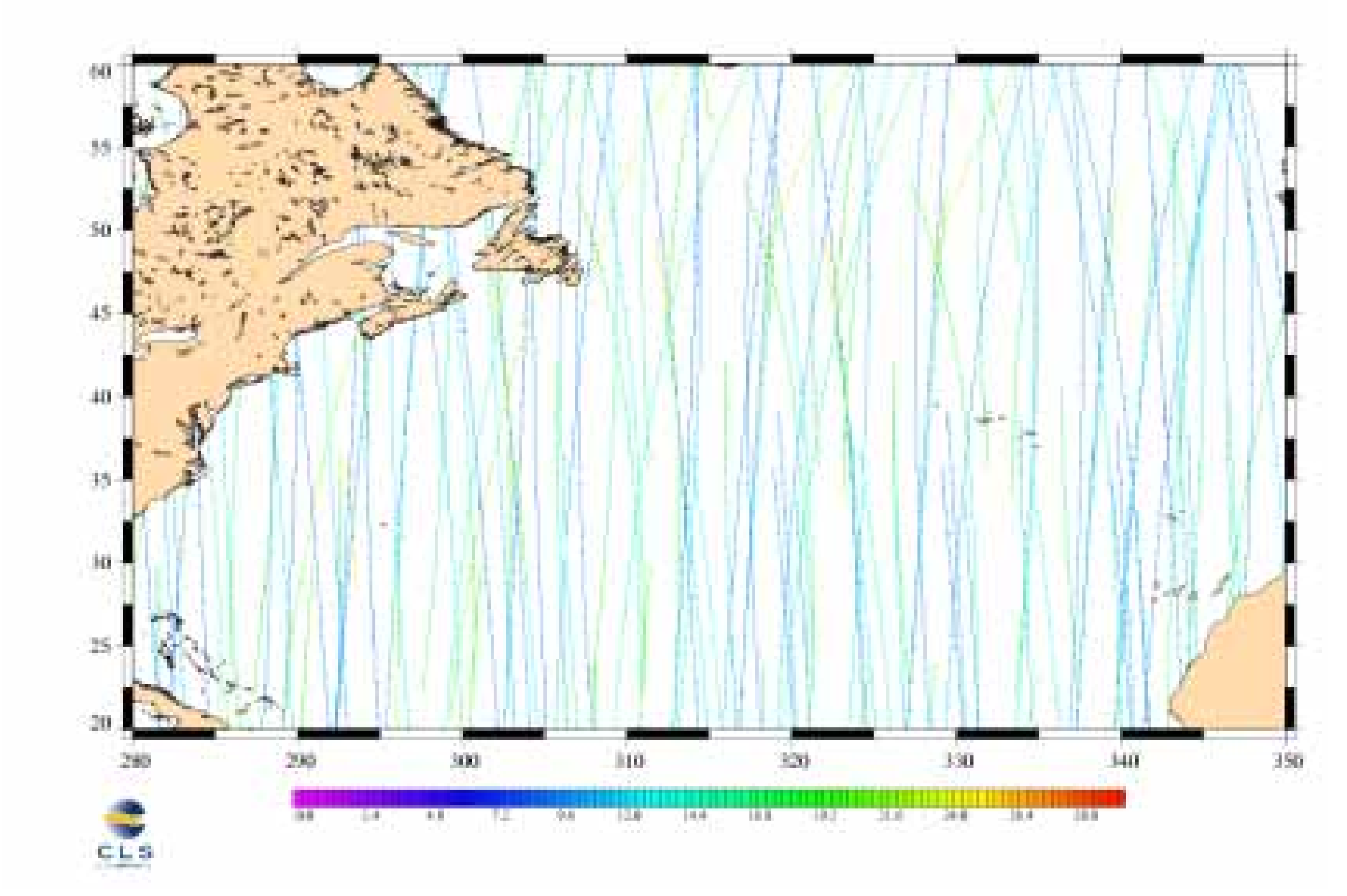

Figure 1 : 3-day sampling of the North Atlantic from one LEO GPS receiver (6 beams) and the September 15, 2001 GPS constellation with incidence angle (a) and errors for 3-second averages associated to the incident angle (b). Units are degree (from 0 to 90) (a) and cm [from 0 to 30] (b).

# 7 Methods

The Los Alamos model (LAM) sea level outputs were first transformed into sea level anomaly data by removing a three-year mean (1993-1995). They were then sub-sampled to obtain simulated GPS signals along the tracks shown on figure 1. A random noise depending on incident angle (see equation 1) was then added to the simulated SLA data. Simulated Jason-1 and ENVISAT data were also obtained by sub-sampling the model fields along TOPEX/POSEIDON and ERS tracks and assuming a 2 cm rms noise for 1-second average.

The simulated data sets were then used to reconstruct the SLA gridded fields using a space-time sub-optimal interpolation method. The method is detailed in Ducet et al. (2000) and Le Traon et al. (2001). It uses the following space (zero crossing of correlation function, ZC) and time (e-folding time, ET) correlation scales :

$ZC = 50 + 250 \left(\frac{900}{Lat^2 + 900}\right)$ km where Lat stands for latitude, in degrees.

ET = 15 days

These scales are intended to represent typical space and time scales of SLA as they can be observed from altimetry. The signal mapping (contrary to the formal error estimation) is not much sensitive, in any case, to these a priori choices when the constraint from the data is strong. The estimations are performed on a regular grid of 1/10° x 1/10°. For a given grid point, all the observations such as $(d/ZC)^2 + (T/ET)^2 < 1$ are taken into account (d is the distance between the observation and the grid point and T is the time difference between the observation and grid point date). Comparison of the reconstructed fields with the reference model fields (resampled on the regular 1/10° x 1/10° grid) allows an estimation of the sea level and velocity mapping error. The velocity errors (and velocity reference fields) were derived from the sea level gradients (finite centred differences) through the geostrophic approximation. In practice, the comparison was made over a 6-month period (1993) with maps calculated every 9 days, i.e. we compared a total of 20 maps. The calculations were done on a large area from 20°N to 60°N and 75°W to 10°W, i.e. covering the full North Atlantic.

In addition to conventional altimeter configurations with one (Jason-1 and ENVISAT) and two satellites (Jason-1+ENVISAT), a GPS only configuration (simulation 1) and a GPS+Jason-1+ENVISAT configuration (simulation 2) were analysed. Simulation 1 allows us to analyze the contribution of GPS altimetry alone while simulation 2 allows us to analyze the complementary of GPS altimetry with the existing and future conventional altimeter missions. It is, indeed, likely that in the coming decade the ocean will be observed at least with two conventional altimeters as it was with TOPEX/POSEIDON and ERS-1/2 and will be soon with Jason-1 and ENVISAT.

To illustrate the methodology, we show on figure 2a the LAM sea level anomaly for a given day. Figures 2b and 2c show the corresponding mapping errors for the GPS and GPS + Jason-1 + ENVISAT configurations. One can note that the GPS configuration is able to capture the main energetic mesoscale signals. In low eddy energy regions, the errors are relatively larger because the measurement noise is much larger than the signal. There, the mapping errors are close to 100 % of the signal variance, i.e. no useful information is brought. Note, however, that as the noise level is taken into account in the analysis, errors are of a few cm rms only, i.e. part of the noise is filtered out in the mapping procedure (when the noise to signal ratio is very large, the estimated field will be close to zero).

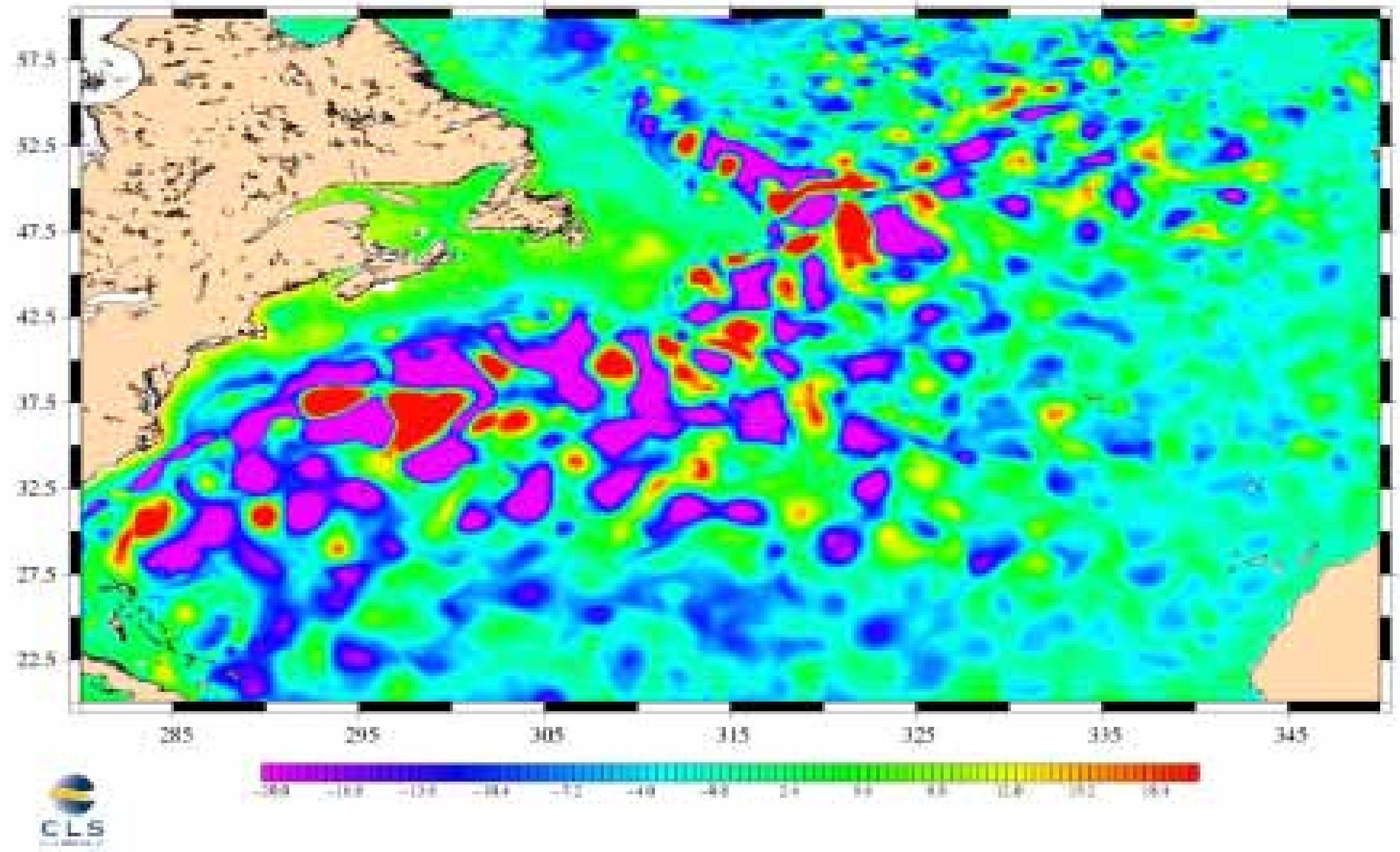

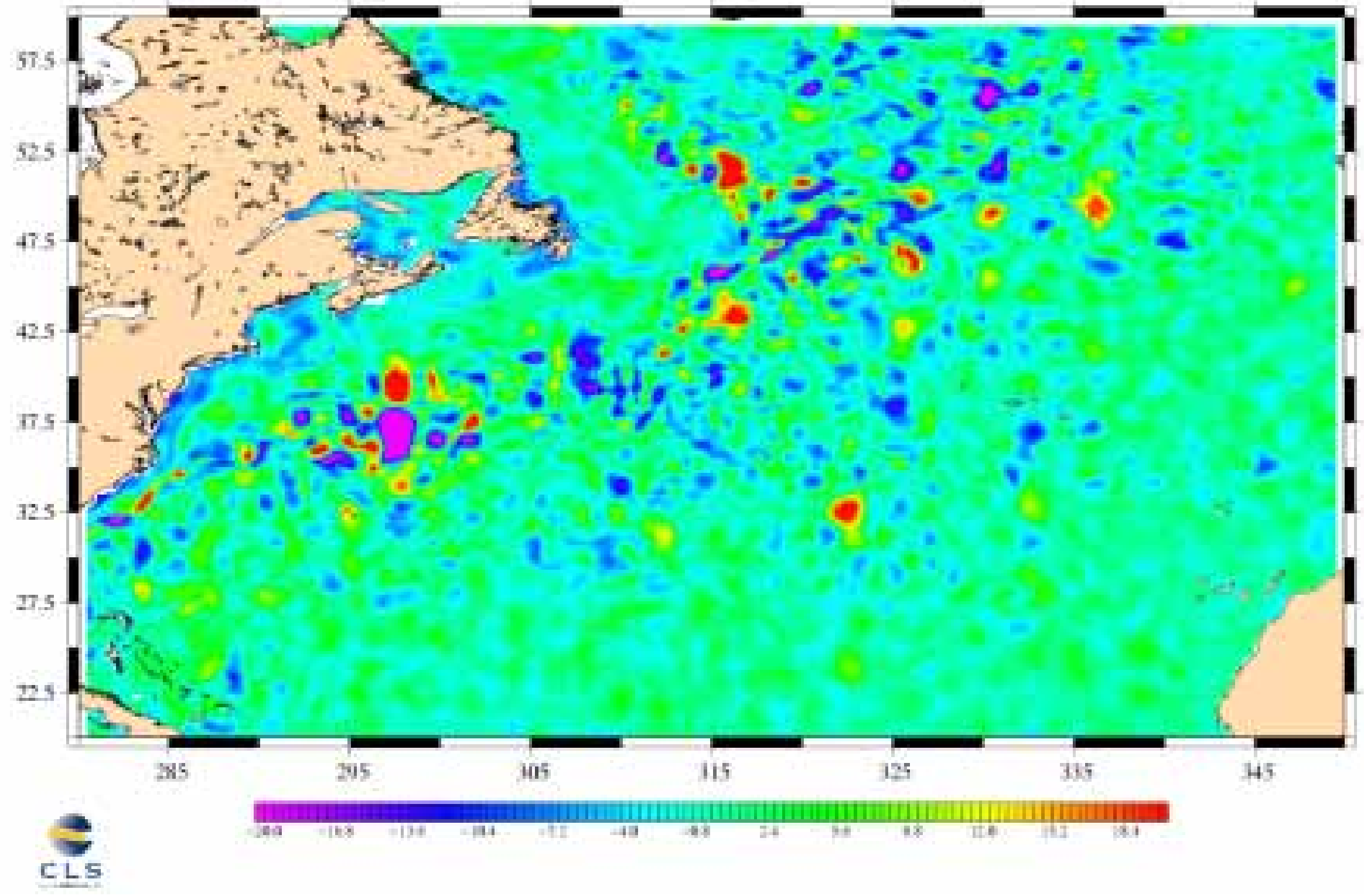

Figure 2a and 2b:  Los Alamos model sea level anomaly field for a given day (a). Corresponding mapping error for the GPS configuration (b). Scale from –20 to 18.4 cm.

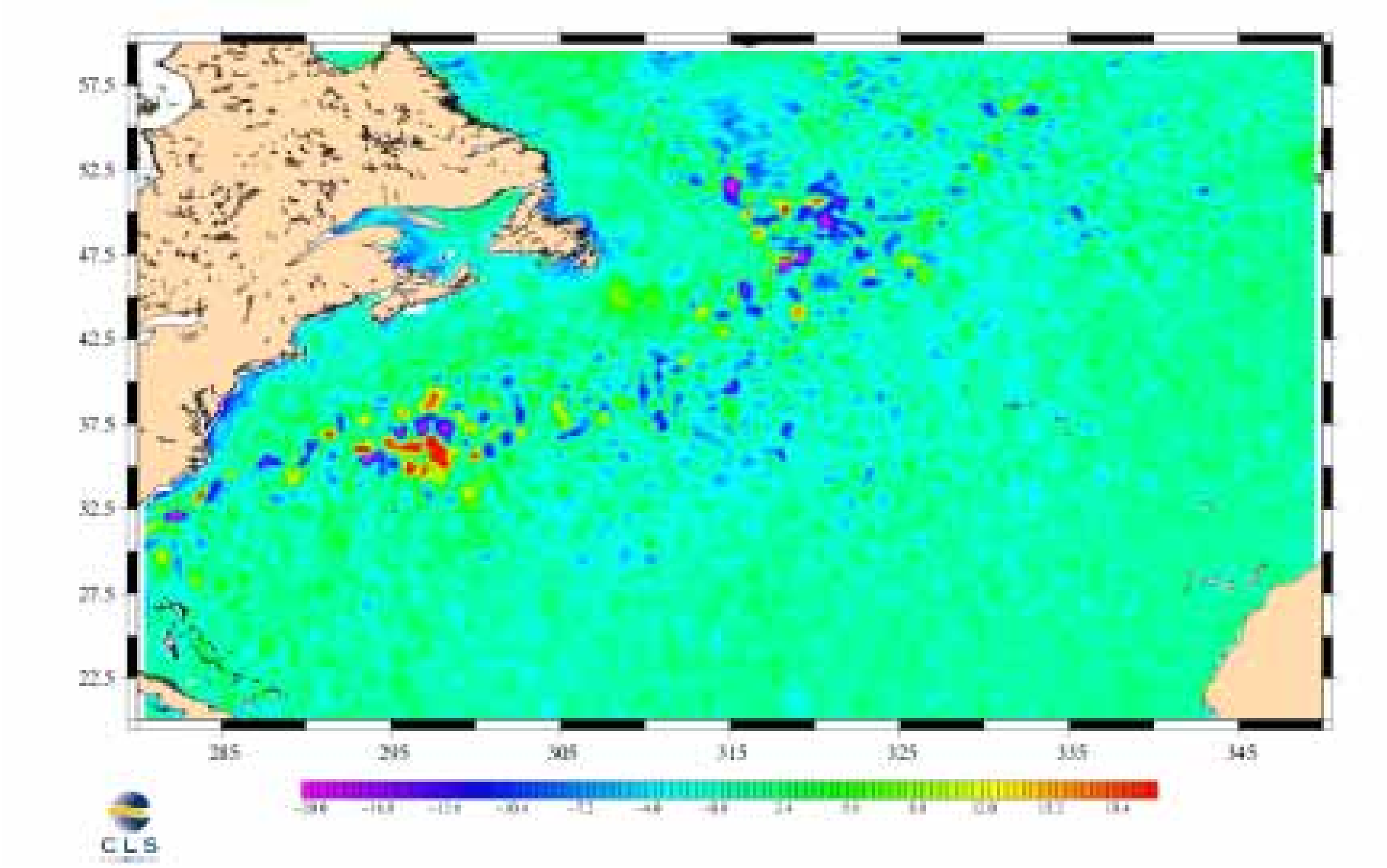

Figure 2c: Corresponding mapping error for the GPS+Jason-1+Envisat configuration.  Units are cm. Scale from –20 to 18.4 cm.

# 8 Results

Mapping errors depend on many factors: sampling characteristics, amplitude, space and time scales of mesoscale ocean signals, relative energy of high frequency signals and measurement noise. All these characteristics vary geographically. As a result, the mapping errors show large and complex geographical variations.  The main effect is, of course, related to the variations in eddy energy. To partly remove this effect, the quadratic relative sea level, zonal and meridional velocity mapping errors (i.e. the ratio of the mapping error variance over the ocean signal variance) were also calculated.

Table 1 summarizes the results obtained for regions where the rms sea level variability is larger than 15 cm rms. This allows us to analyze the mapping capabilities for regions whose dynamics are dominated by mesoscale variability. These are the regions where we expect a significant contribution of GPS altimetry.

| | H | U | V |
|---|---|---|---|
| **Jason-1** | **22.1** | **44.1** | **55.0** |
| **ENVISAT** | **12.1** | **34.5** | **45.1** |
| **Jason-1+ENVISAT** | **8.4** | **26.6** | **33.8** |
| **GPS** | **12.6** | **31.1** | **39.9** |
| **GPS+Jason-1+ENVISAT** | **4.7** | **20.8** | **26.4** |

Table 1: Sea Level (H), zonal (U) and meridional (V) velocity, all in (%).

These results show that in regions with large mesoscale variability GPS is performing better than Jason-1 and almost as well as ENVISAT. The combination of GPS with Jason-1 and ENVISAT should also improve the sea level mapping by a factor of about two and the velocity mapping by almost 33%.

This is mainly due to the dense sampling (and, in particular, the high frequency sampling) offered by GPS. In the simulated sampling we used, there are about 6 measurement points every second, i.e. 6 times more than for a conventional altimeter. The GPS measurement noise variance is, of course, much larger than for a conventional altimeter (by a factor of up to 60) but, in high eddy variability regions, it remains smaller than the signal variance. An observation with a 10 cm rms measurement noise yields a relative mapping error of about 10% in the Gulf Stream area (where the rms sea level variability is larger than 30 cm). This is smaller than the mapping error obtained from ENVISAT and equivalent to the mapping error obtained from the combination of Jason-1 and ENVISAT.

Figures 3 and 4 show the rms sea level mapping errors for simulations 1 and 2 respectively. Figure 3 is clearly related to the sampling characteristics of GPS (see figure 1a). Errors are large when there is a poor sampling (e.g. near 63°W and between 30°N and 40°N). They are also relatively large in low eddy energy regions. This is much improved when GPS data are merged with Jason-1 and ENVISAT.

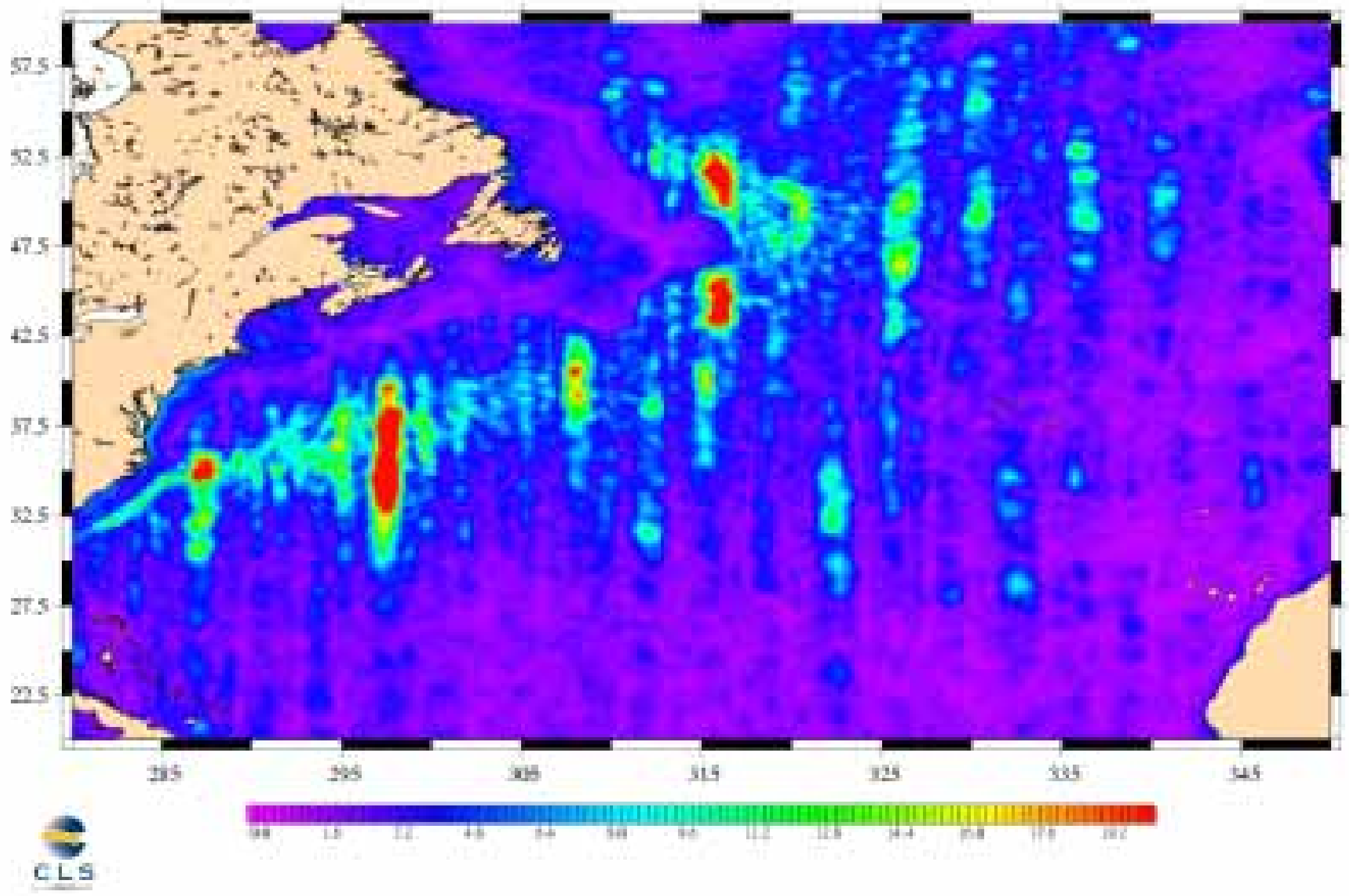

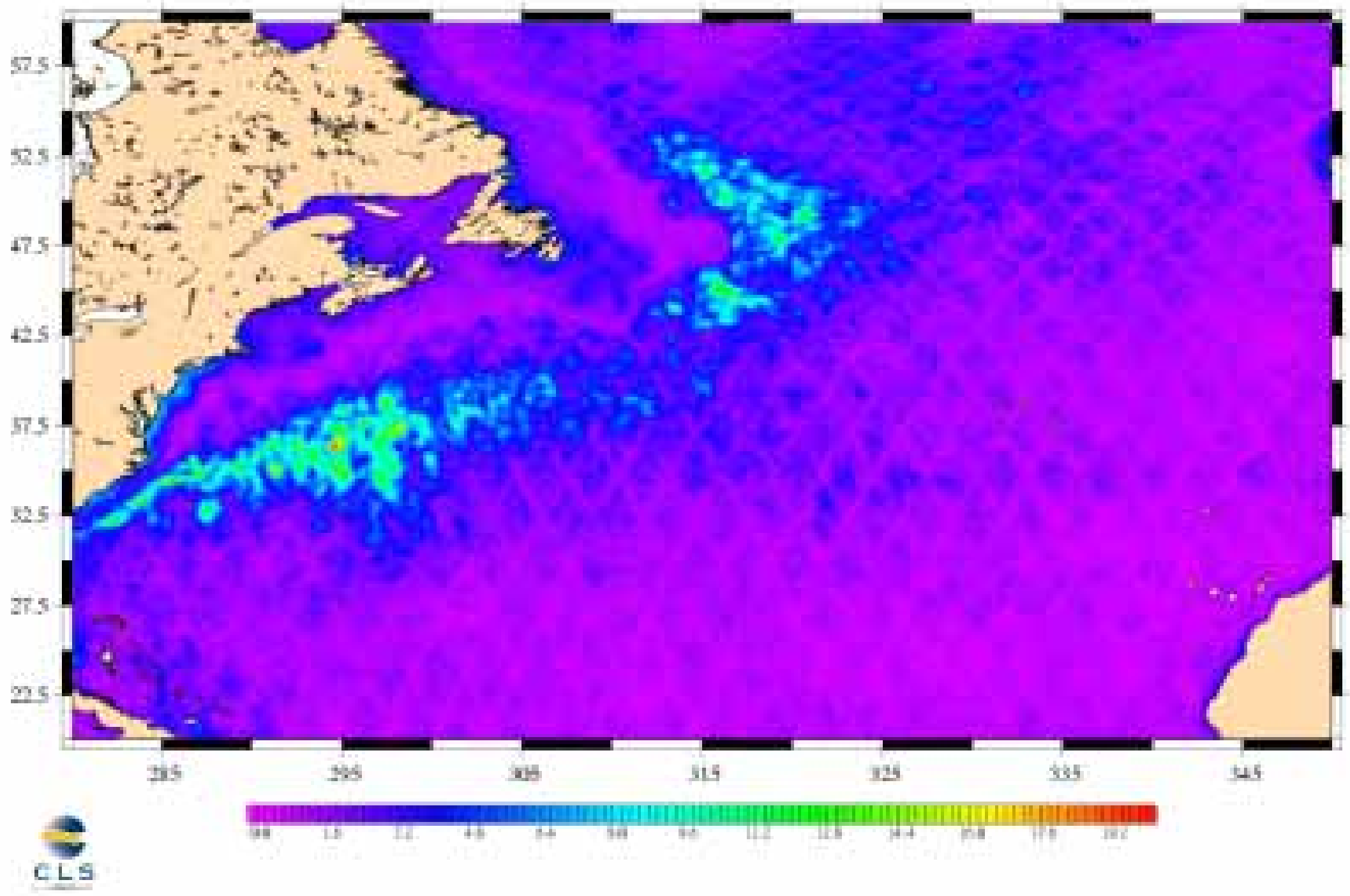

<u>Figures 3 and 4</u>: Rms sea level mapping error for GPS (upper figure), GPS+Jason-1+ ENVISAT (lower figure). Scale from –20 to 18.4 cm.

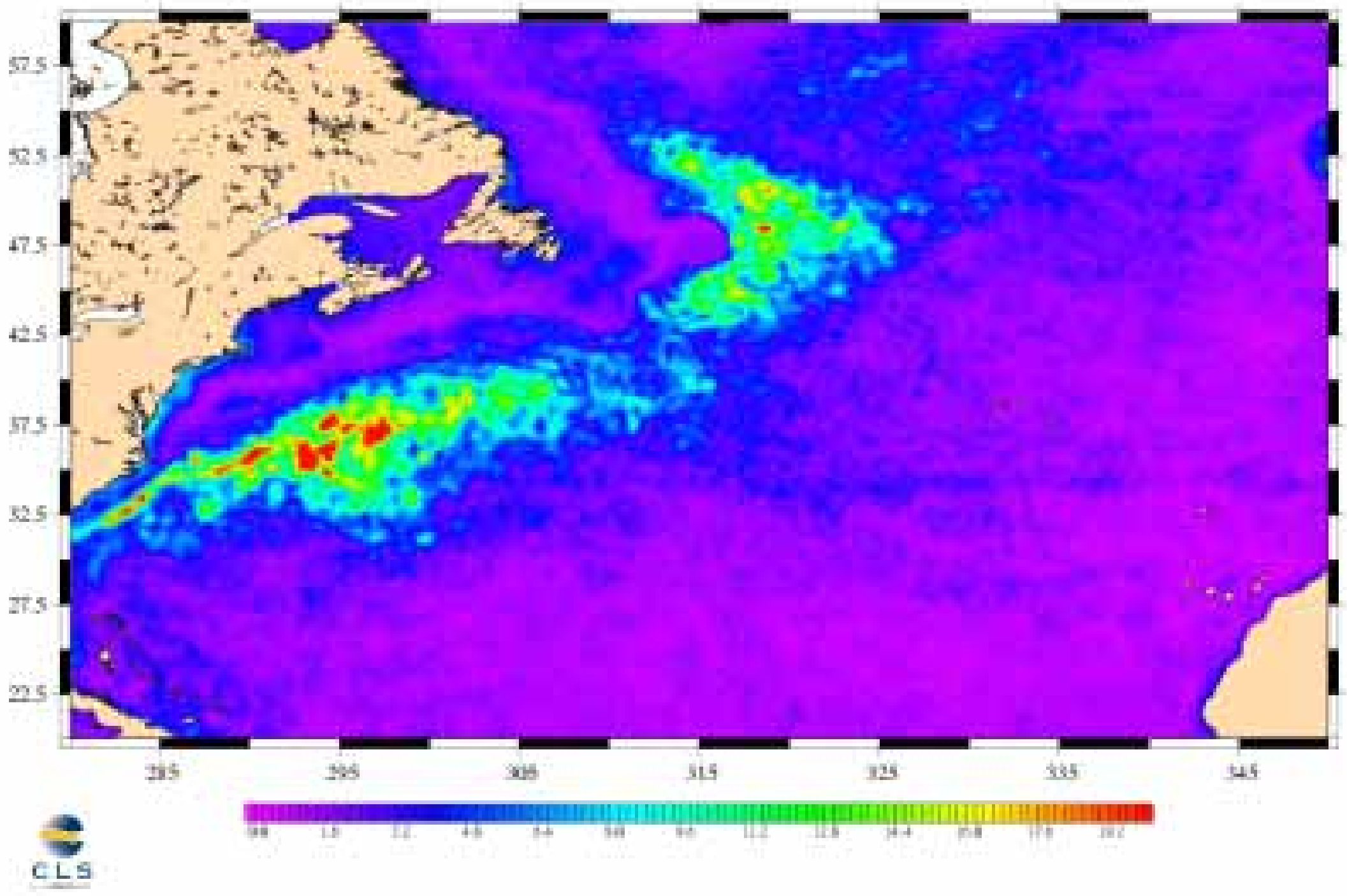

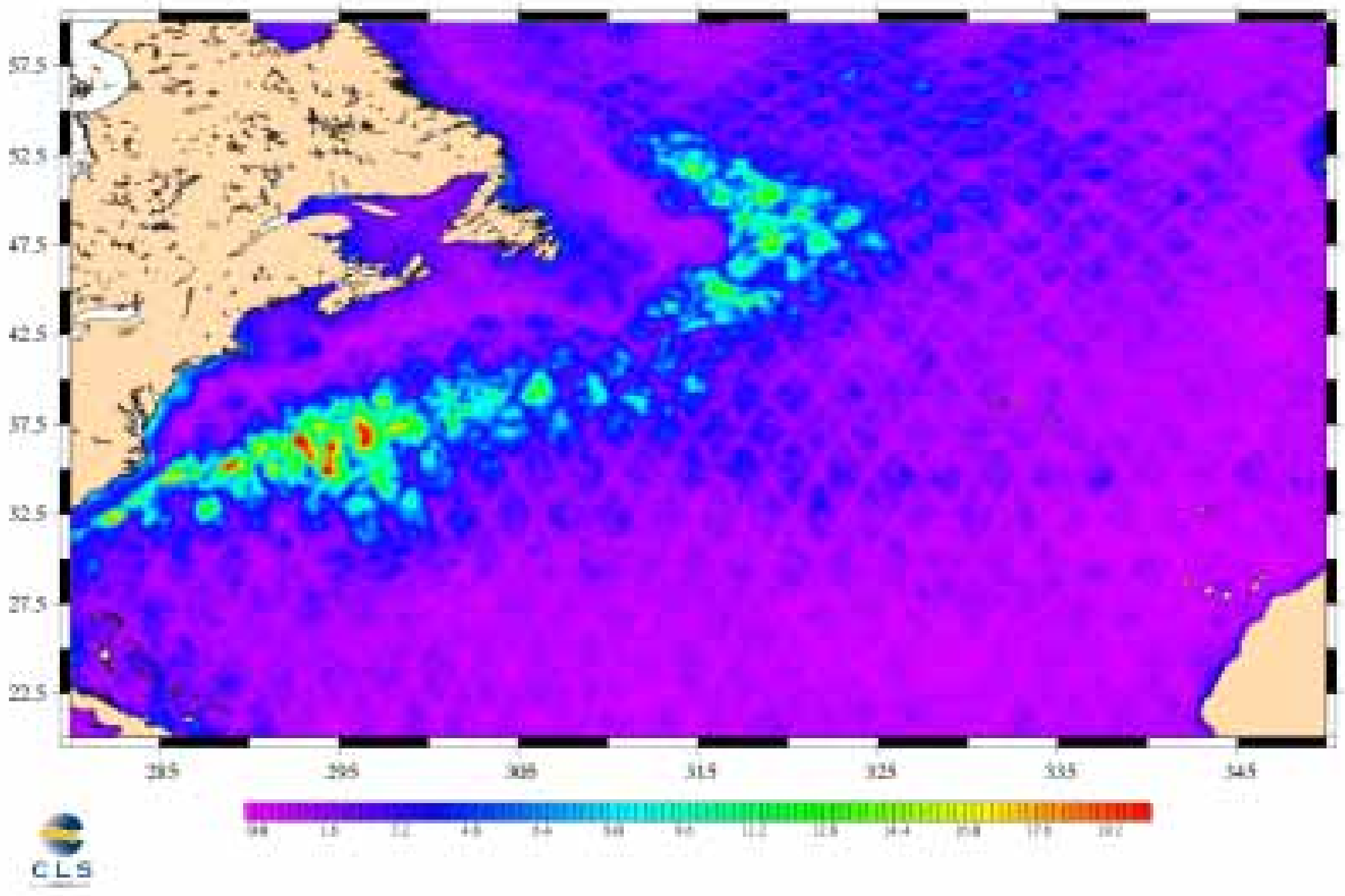

<u>Figures 5 and 6</u> : Rms sea level mapping error for ENVISAT (upper figure) and Jason-1+ ENVISAT (lower figure). Units are cm.

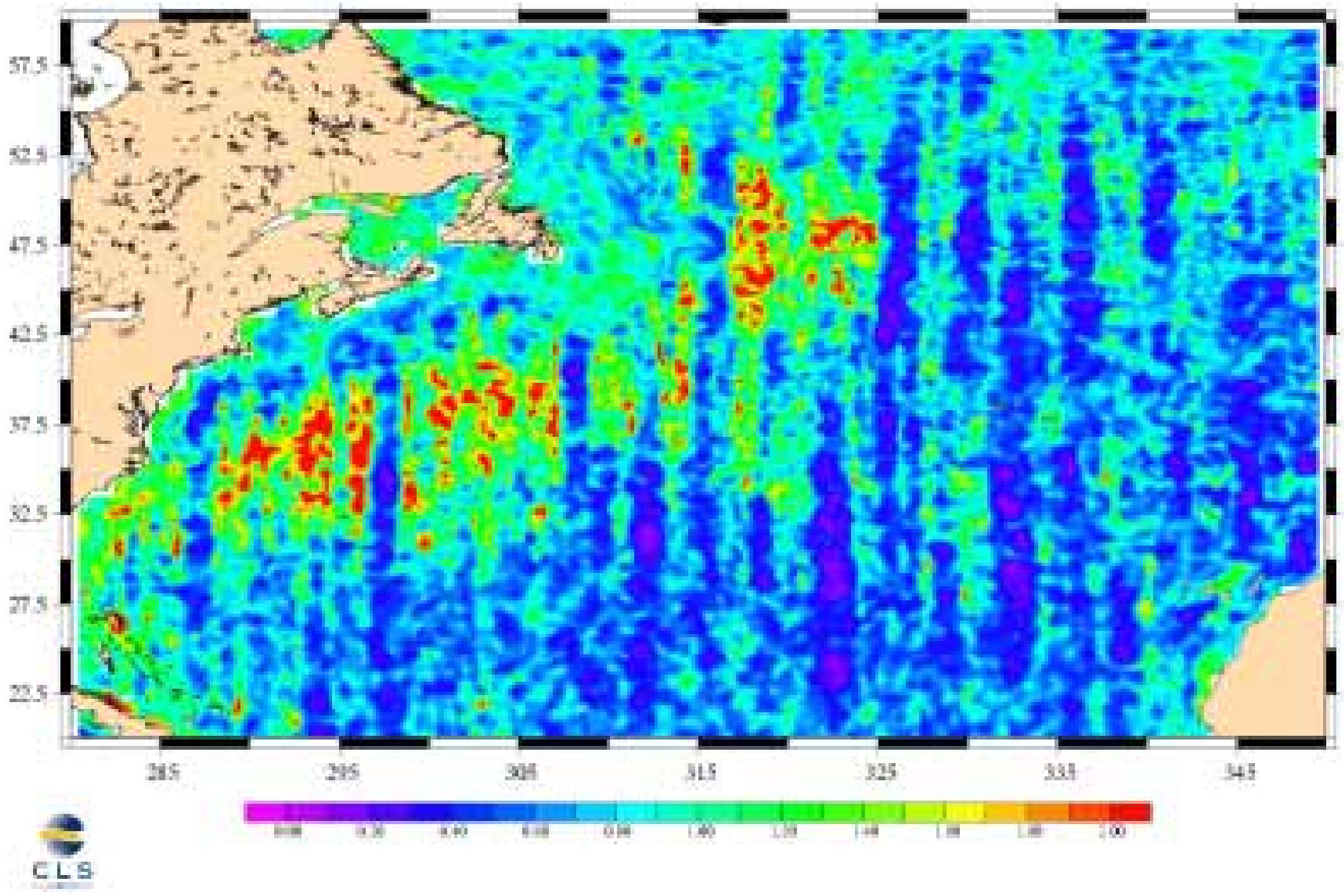

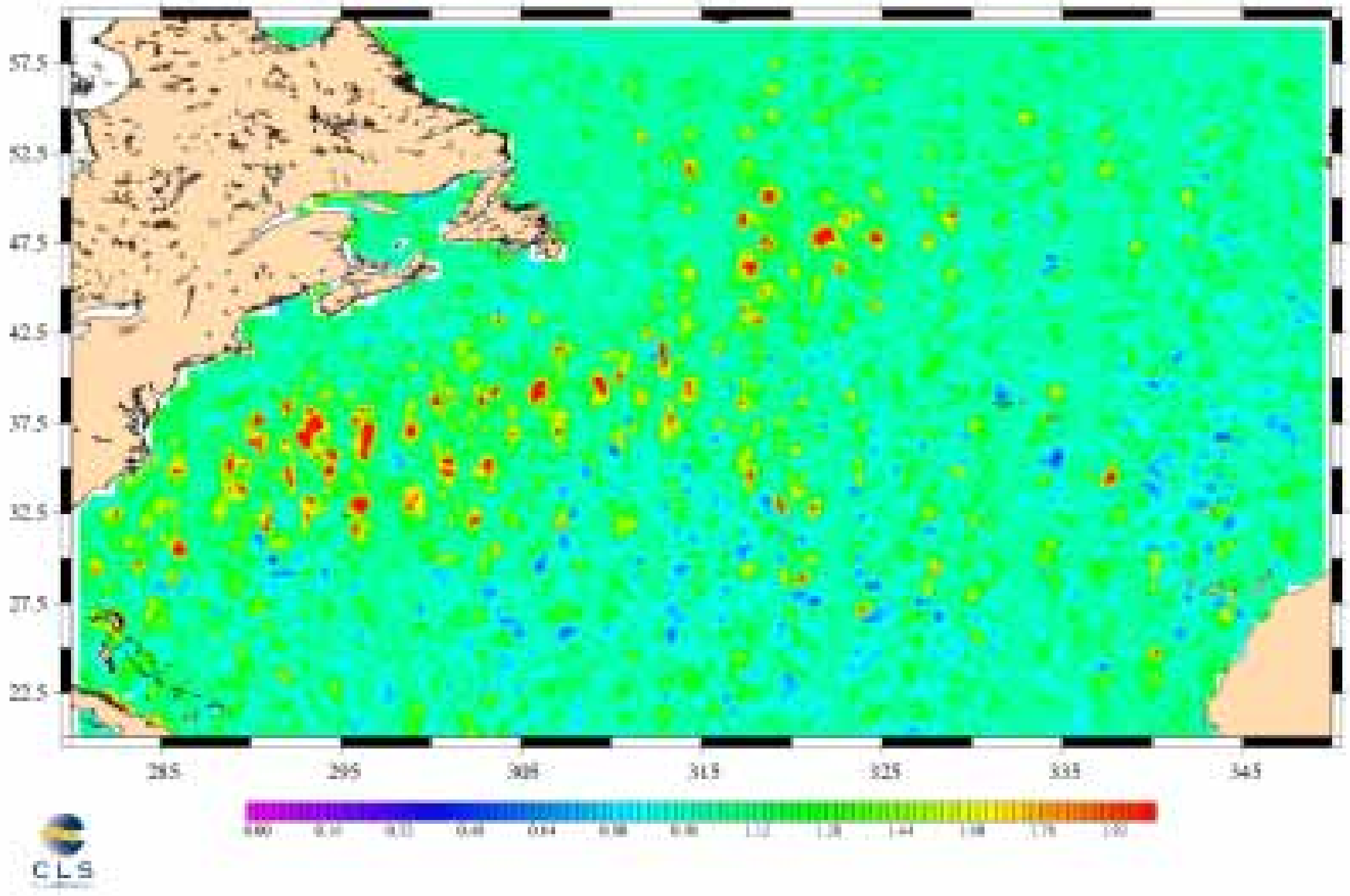

Figures 7 and 8 : rms ENVISAT sea level mapping error over rms GPS sea level mapping error (upper figure, scale from 0 to 2.1). Rms Jason-1+ENVISAT sea level mapping error over rms GPS+Jason-1 + ENVISAT sea level mapping error (lower figure, scale from 0 to 2.0).

Figures 3 and 4 can be compared with sea level mapping errors from one altimeter (ERS or ENVISAT) (figure 5) and from the combination of two conventional altimeters (T/P and ERS or Jason-1 and ENVISAT) (figure 6). As in table 1, these results illustrate the contribution of GPS altimetry. In large eddy variability regions and well-sampled regions, GPS is performing much better than Jason-1 and ENVISAT alone and as well as Jason-1+ENVISAT.

To quantify, the contribution of GPS altimetry, the ratio between the ENVISAT and GPS mapping error variance (figure 7) and the Jason-1+ENVISAT and GPS+Jason-1+ENVISAT (figure 8) were also calculated. Compared to ENVISAT, in the Gulf Stream region and in well-sampled regions, the GPS mapping error variance are divided by a factor of up to 4. Compared to Jason-1+ENVISAT, the combination of GPS, Jason-1 and ENVISAT should also allow a reduction of the mapping error variance by a factor of up to four (mainly in between the Jason-1 tracks where the mapping from the combination of Jason-1 and ENVISAT is less accurate and in regions well sampled by GPS).

# 9 Formal mapping errors

The formal mapping errors (expressed in percentage of signal variance) were also derived from these analyses. These errors only depend on sampling, noise and signal characteristics and do not depend on model fields. They provide a robust estimation of the mapping capabilities that only depend on our choice of signal covariance functions. Figures 9 and 10 show the sea level formal mapping errors for simulations 1 and 2 respectively. The ratio between the ENVISAT and GPS formal mapping error variance (figure 11) and the Jason-1 +ENVISAT and GPS+Jason-1+ENVISAT (figure 12) are also shown. Results are consistent with those discussed in section 4. In high eddy energy regions and well sampled regions, GPS and GPS+Jason-1 +ENVISAT are generally performing much better than ENVISAT and Jason-1+ENVISAT respectively (mapping error variances are divided by a factor of up to 4).

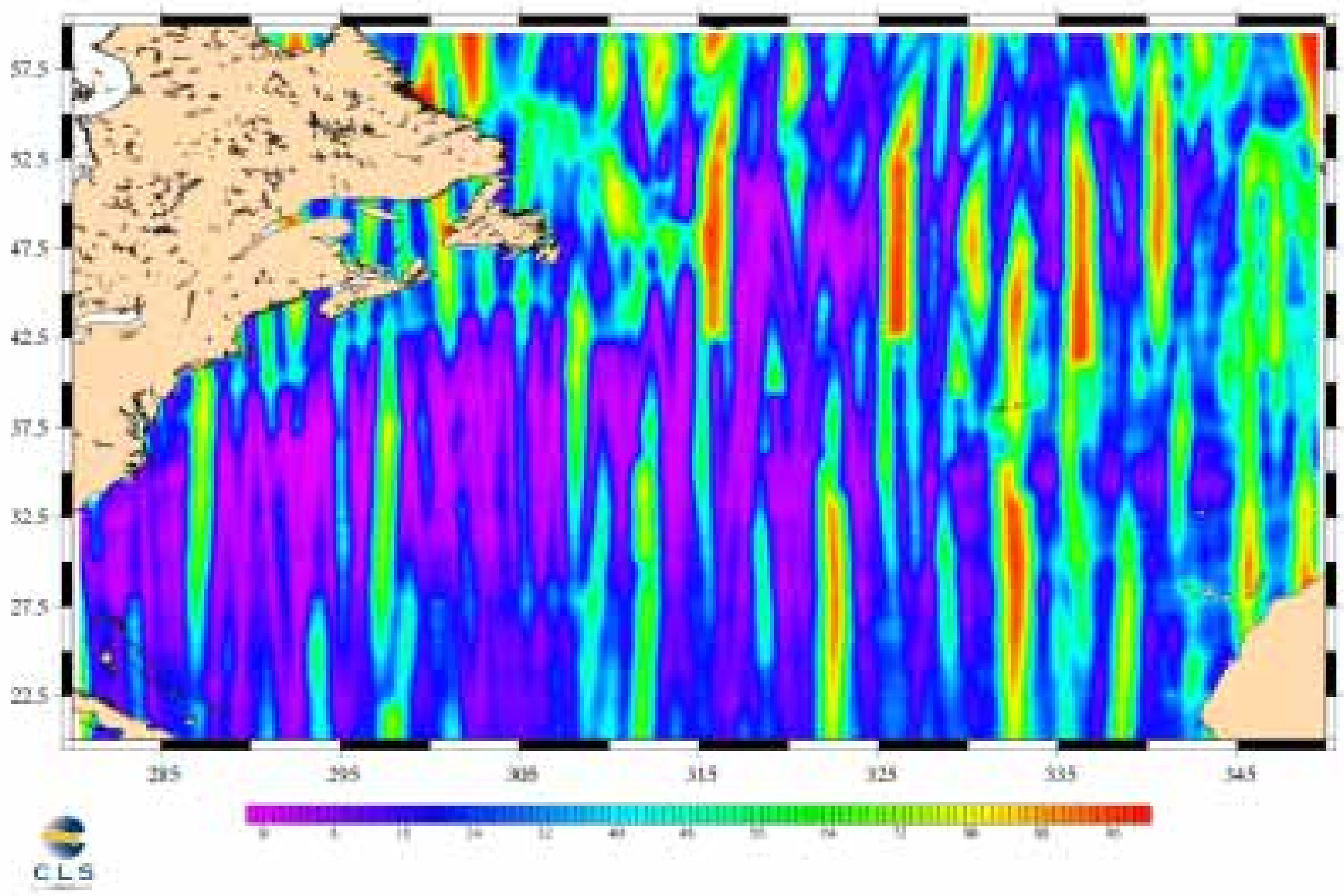

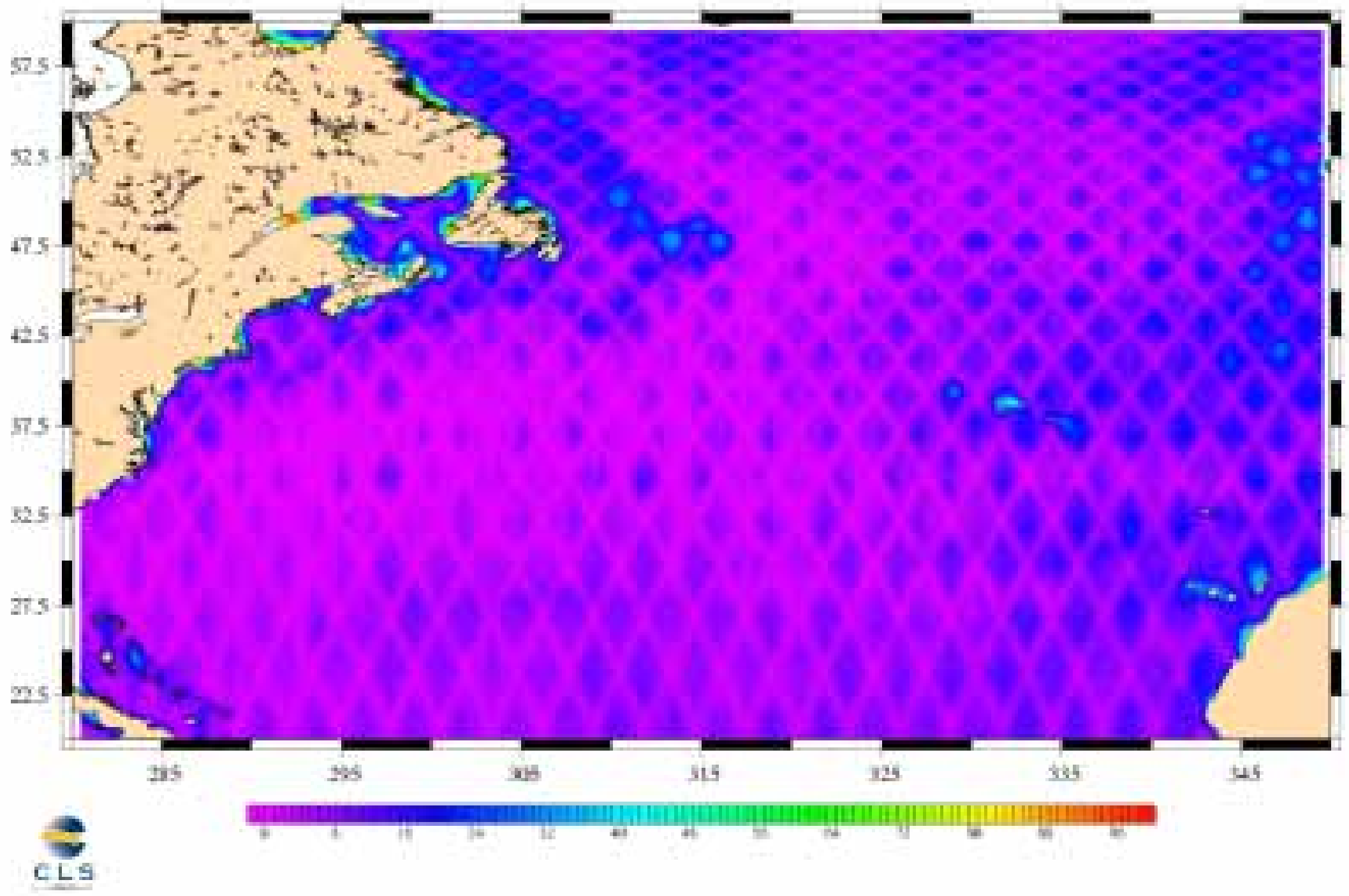

Figures 9 and 10: Formal sea level mapping error for GPS (upper figure), and GPS+Jason-1 + ENVISAT (lower figure). Units are percentage of signal variance, scale from 0 to 100.

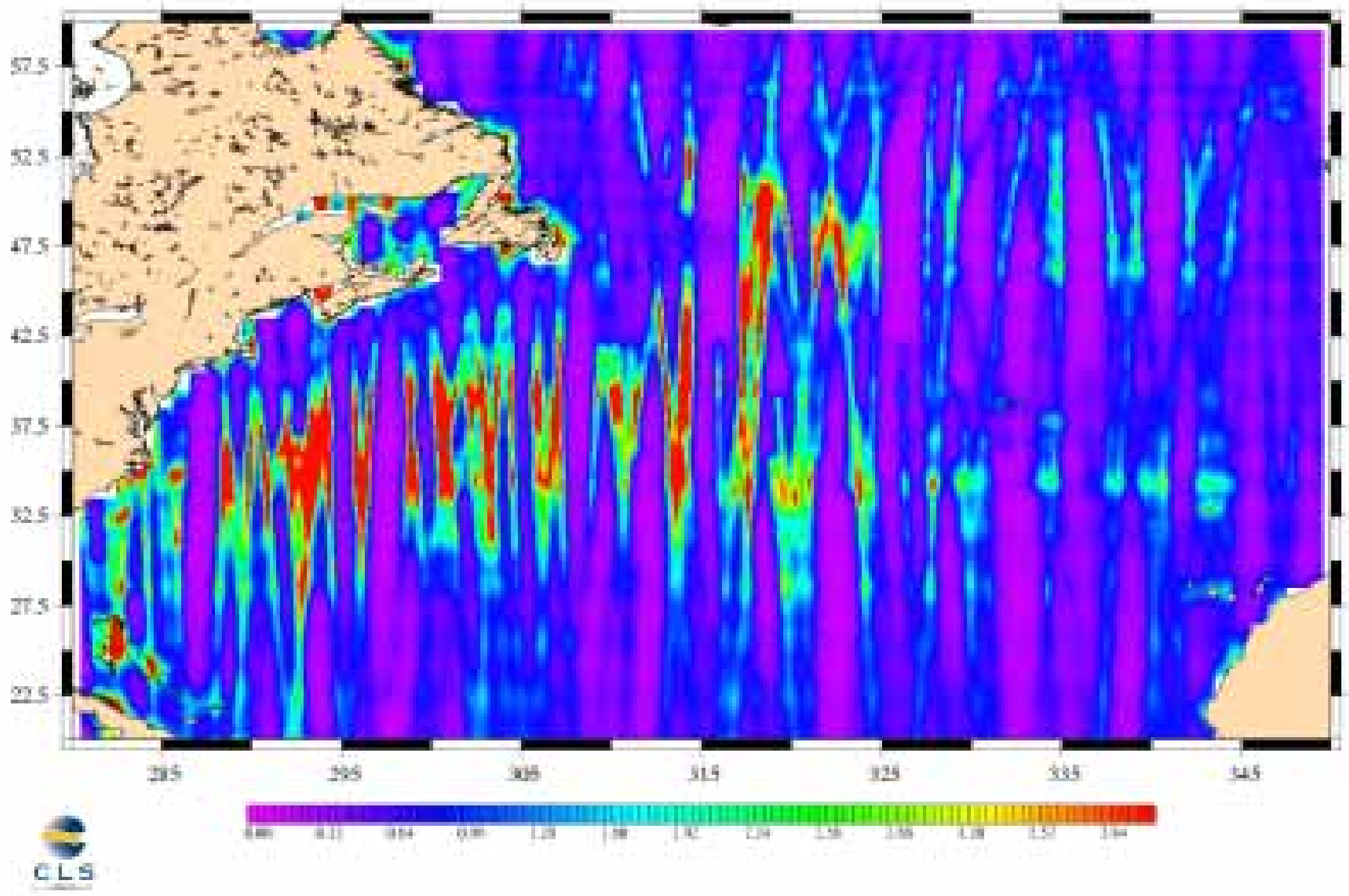

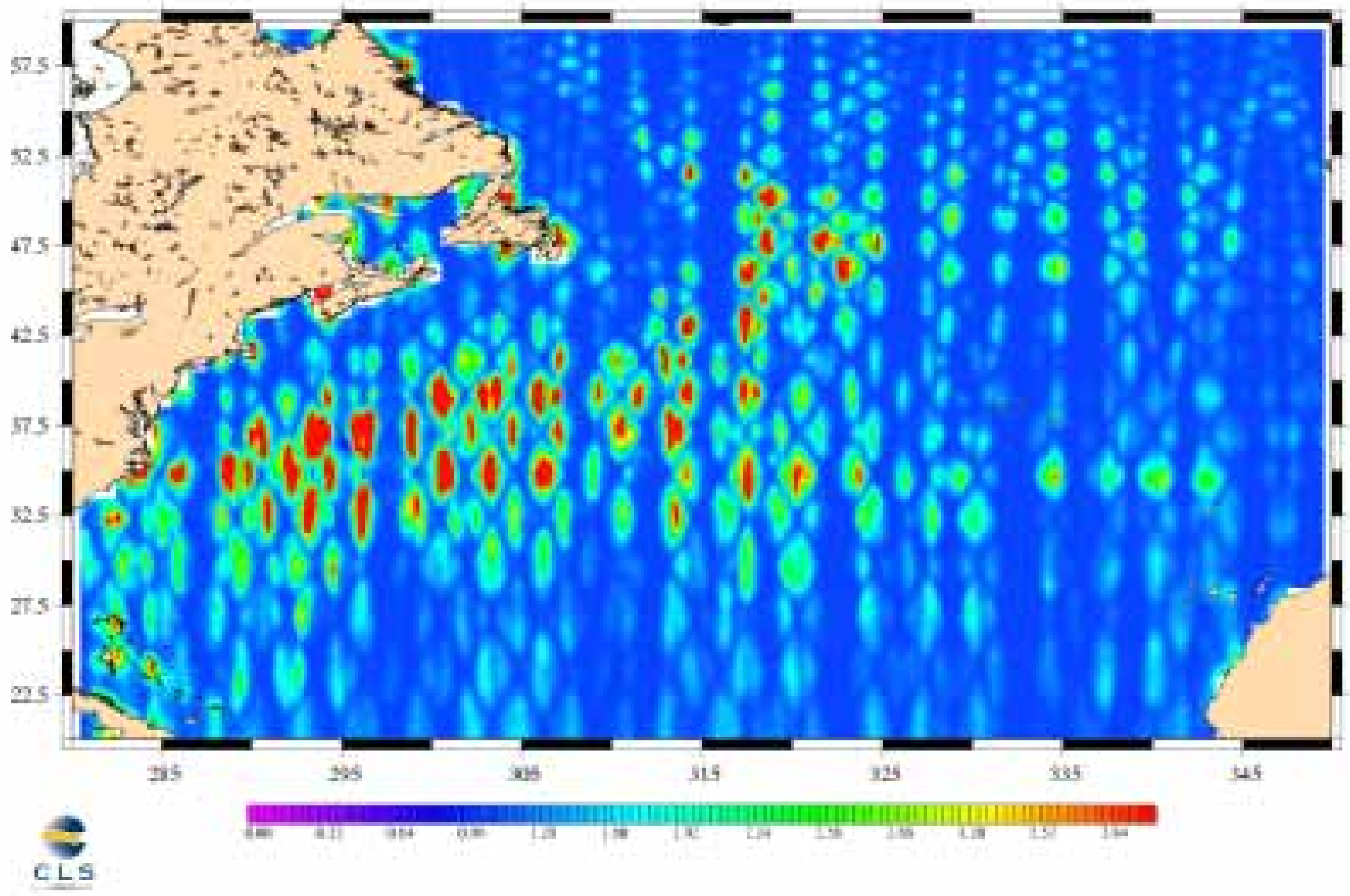

Figures 11 and 12 : ENVISAT formal mapping error variance over GPS formal mapping error variance (upper figure). Jason-1+ENVISAT formal mapping error variance over GPS+Jason-1+ENVISAT formal mapping error variance (lower figure). Scale from 0 to 4.

# 10 Conclusions

As we have seen, a single PETREL-like mission (with a multi-beam capability) should have a very significant impact on the mapping of the mesoscale variability. According to the chosen error budget, it should allow us the map the mesoscale variability in high eddy variability regions almost as well as ENVISAT. In well-sampled regions, the mapping should be as well as the one derived from the combination of Jason-1 and ENVISAT.

Moreover, the combination of GPS with Jason-1 and ENVISAT will improve the sea level mapping derived from the combination of Jason-1 and ENVISAT by a factor of about 2. In well-sampled regions, the improvement could reach up to a factor of 4.

Although there are other sources of errors that should be taken into account (e.g. propagation effects, orbit, geoid), this is already a very encouraging result. The dense and high frequency sampling offered by multi-beam GPS altimetry is thus likely to compensate for the large noise level (15 to 20 cm for 1 second-average) (in large eddy variability regions).

Note finally that the study only considered GPS satellites and a six-beam system; results should significantly improve when both GPS and Galileo satellites are used and when more beams are used.